\documentclass[a4paper,fleqn,usenatbib]{mnras}
 
\usepackage[T1]{fontenc}
\usepackage{ae,aecompl}

\usepackage{graphicx}	
\usepackage{amsmath}	
\usepackage{amssymb}	

\def \gsim { \lower .75ex \hbox{$\sim$} \llap{\raise .27ex \hbox{$>$}} }
\def \lsim { \lower .75ex \hbox{$\sim$} \llap{\raise .27ex \hbox{$<$}} }

\newcommand{\civ}{C\,{\sc iv}\ }

\newcommand{\mgii}{Mg\,{\sc ii}\ }

\newcommand{\oiii}{[O\,{\sc iii}]\ }

\title[Changing-Look Quasars in SDSS]{A Systematic Search for Changing-Look Quasars in SDSS}

\author[MacLeod et al.]
{Chelsea L. MacLeod$^{1}$\thanks{email: cmacleod@roe.ac.uk}, Nicholas P. Ross$^{1}$, Andy Lawrence$^{1}$, Mike Goad$^{2}$, \newauthor Keith Horne$^{3}$, William Burgett$^{4}$, Ken C. Chambers$^{5}$, Heather Flewelling$^{5}$, \newauthor Klaus Hodapp$^{5}$,  Nick Kaiser$^{5}$, Eugene Magnier$^{5}$, Richard Wainscoat$^{5}$,  \newauthor Christopher Waters$^{5}$\\
$^{1}$Institute for Astronomy, University of Edinburgh, Royal Observatory, Edinburgh, EH9 3HJ, U.\ K.\\
$^{2}$Department of Physics and Astronomy, University of Leicester, University Road, Leicester, LE1 7RH, U.\ K.\\
$^{3}$SUPA Physics \& Astronomy, North Haugh, St Andrews KY16 9SS, U.\ K.\\
$^{4}$GMTO Corp., 251 S.\ Lake Ave., Pasadena, CA 91101 USA\\
$^{5}$Institute for Astronomy, University of Hawaii, 2680 Woodlawn Dr., Honolulu, HI 96822, USA
}

\date{Accepted XXX. Received YYY; in original form ZZZ}

\pubyear{2015}

\begin{document}
\label{firstpage}
\pagerange{\pageref{firstpage}--\pageref{lastpage}}
\maketitle

\begin{abstract}
We present a systematic search for changing-look quasars based on repeat photometry from SDSS and Pan-STARRS1,
along with repeat spectra from SDSS and SDSS-III BOSS. Objects with
large, $|\Delta g|>1$~mag photometric variations in their light curves
are selected as candidates to look for changes in broad emission line
(BEL) features. Out of a sample of 1011 objects that satisfy our
selection criteria and have more than one epoch of spectroscopy, we find 10
examples of quasars that have variable and/or ``changing-look'' BEL
features. Four of our objects have emerging BELs; five have
disappearing BELs, and one object shows tentative evidence for having
both emerging {\it and} disappearing BELs. With redshifts in the range
$0.20<z<0.63$, this sample includes the highest-redshift changing-look
quasars discovered to date. We highlight the quasar
J102152.34+464515.6 at $z=0.204$. Here, not only have the Balmer
emission lines strongly diminished in prominence, including H$\beta$
all but disappearing, but the blue continuum $f_{\nu}\propto
\nu^{1/3}$ typical of an AGN is also significantly diminished in the
second epoch of spectroscopy. Using our selection criteria,
we estimate that $>15$\% of strongly variable luminous quasars display changing-look BEL features on rest-frame timescales of 8 to 10 years. 
Plausible timescales for variable dust extinction are factors of 2--10 too long to explain the dimming and 
brightening in these sources, and simple dust reddening models cannot reproduce the BEL changes.  On the other hand, an advancement such as disk reprocessing is needed if the observed variations are due to accretion rate changes.
\end{abstract}




\section{Introduction}
Due to modern photometric, spectroscopic, and time-domain sky surveys,
we are now able to distinguish between the ubiquitous
properties of quasars and rare, discrepant behavior, suggestive of new
physics.  The continuum variability of a
quasar\footnote{We use the term quasar to mean a
bolometrically luminous, $\gtrsim10^{46}$ erg s$^{-1}$, active
galactic nucleus (AGN).} is typically 0.2~mag on timescales of
$\sim$months to years and has been well-characterized for large quasar
samples using the Sloan Digital Sky Survey \citep[SDSS;
e.g.,][]{van04,wil05,ses07,schm10,but11,mac12} and more recently
Pan-STARRS1 \citep[PS1;][]{mor14,mor15,sim15}.  A common interpretation of
this variability involves localized accretion disk instabilities,
but the precise mechanisms are still under debate \citep[e.g.,][and
references therein]{dex11,kel11,kok15}.  
For example, the color dependence and near simultaneity of
variability pose severe problems \citep[see][and references
therein]{law12}.  In Seyfert galaxies, time lags of hours to days suggest that the
optical continuum variability is driven at least in part by
reprocessing of EUV or X-ray light
\citep[e.g.,][]{col98,ser05,cac07,sha14,ede15} but this may be due to inner
cloud reprocessing rather than the disc itself \citep{law12}, and the
reliability of the interband lags has been questioned \citep{kor01,gas07}.

While the UV/optical continuum varies ubiquitously among quasars,
typically in a `bluer-when-brighter' fashion
\citep[e.g.,][]{wil05,sch12,rua14}, the broad emission lines (BELs)
seen in quasar spectra are usually less variable and lagged with
respect to the continuum, as seen in reverberation mapping studies
\citep[e.g.,][]{cla91, pet93, gri12, pet14, shen15}. Results from such
studies indicate that the BELs are formed by photoionization by
continuum photons, and that the broad line region (BLR) spans a range
in density, ionization state, and distance from the central black
hole, which is typically 
$R/c \approx 0.1{\rm d} (L/10^{46} {\rm erg~s}^{-1})^{1/2}$. Some BELs show a 
Baldwin effect, where the line equivalent width (EW) decreases with 
increasing continuum luminosity \citep{bal78,kin90}.  In general, it 
seems that the underlying spectral energy distribution (SED) influences 
the wind structure and the high-ionization BEL parameters \citep{ric11}.
Because H$\beta$ in particular, and to some extent \mgii basically count 
the number of ionizing photons and are less sensitive to changes in the 
AGN SED than the high ionization lines, neither line shows a particularly 
strong global Baldwin effect \citep{die02}. However, an intrinsic Baldwin 
effect is seen and expected in an individual source 
\citep[for H$\beta$ see][]{gil03,goa04,cac06}.
Furthermore, BELs may respond differently to changes in 
ionizing continuum flux if the average formation radius is
different. For example, \citet{goa93} and \citet{obr95} show that
\mgii responds less to continuum changes relative to the other BELs
(e.g., H$\beta$) due to its larger average formation radius 
and weak intrinsic response.

Emerging or disappearing BELs have been discovered in several
relatively local AGN
\citep{kha71,toh76,pen84,coh86,bis99,are99,era01,sha14,den14,li15},
and recently in a higher-redshift $z=0.3$ quasar
SDSS~J015957.64+003310.4 \citep[hereafter, J0159+0033;][]{lam15}.
These BEL changes are often associated with large, factor $\gtrsim 10$
changes in the continuum flux; the two are probably linked to the same
physical mechanism, as such BEL and continuum changes are rare in
AGN. This ``changing-look'' behavior is valuable for understanding the
structure of the accretion disk and BLR.  One possible physical scenario that
could explain this behavior is variable obscuration by dust or gas
clouds passing across the line of sight
\citep[e.g.,][]{goo89,tra92,ris09}.  While strong arguments exist for
variable absorption of the X-ray emitting region in some AGN
\citep{ris09}, it is relatively difficult to explain variable
obscuration of the much larger UV/continuum region, as it is unclear
what population of clouds would exist at sufficiently large radii
\citep[e.g.,][]{nen08a, nen08b}.  Indeed, a simple change in extinction
fails to explain the observations of J0159+0033
\citep{lam15}. Instead, the transition might be due to a change in
available ionizing flux from the central engine. \citet{lam15} suggest
that such a change would be consistent with the disk-wind scenario of
\citet{eli14}, where a decrease in the accretion rate, and
consequently the level of ionizing flux, would cause the BELs to
disappear as the system evolves from Type 1 to 1.2/1.5 to
1.8/1.9\footnote{We follow the classical definitions of optical types,
where the total flux in H$\beta$ relative to [OIII] decreases going
from type 1.2 to 1.8, and type 1.9 is defined as having broad
H$\alpha$ but lacking any broad H$\beta$ \citep{ost81}.}. An alternative 
scenario was proposed by  \citet{mer15} in which  J0159+0033 underwent a 
flaring episode due to a tidal disruption event of a star by the supermassive black hole.

Motivated by the discovery of a single changing-look quasar in SDSS by
\citet{lam15}, in this paper we undertake a systematic search for
similar objects with candidate `changing-look' quasars being selected
from their light curves over the course of SDSS and PS1. Our search is
sensitive to emerging or disappearing BELs that may be associated
with strong increases or decreases in flux in quasars which were more
luminous than $M_i = -22.0$ with $i>15.0$ roughly ten years ago.
In particular, what is being observed and reported here is the (dis)appearance 
of broad Balmer emission lines, and a strengthening (weakening) of MgII emission. 

The outline of the paper is as follows. In Section~\ref{data}, we
describe the SDSS and PS1 data used in this study.  In
Section~\ref{sample}, we describe the sample selection.  We present
the results from our search in Section~\ref{results}, describing a set
of 10 quasars, five of which have disappearing BELs, four of which
have appearing BELs, and one object that potentially has evidence for
both disappearing and appearing BEL behavior.  In Section~\ref{disc},
we discuss the timescales associated with the changing-look BEL
phenomenon, and calculate probabilities of observing the quasar BEL
changes as a function of (rest-frame) timescale.  We conclude in
Section~\ref{conc}. All photometric measurements are normalized to the
AB-magnitude system. Where needed, we adopt a flat $\Lambda$CDM
cosmology with $H_{0} = 70$ km s$^{-1}$ Mpc$^{-1}$ and
$\Omega_{\rm m} = 0.30$.

\section{Data}\label{data}
For our search of candidate changing-look quasars, we utilize three
imaging data sets as well as spectroscopy from the SDSS-I/II and
SDSS-III surveys.

\subsection{Imaging Data}
We use imaging data from the SDSS, SDSS-III, PS1 and Catalina Sky
Survey.  As a guide to the baseline of the observations, SDSS started
its imaging campaign in 2000 and concluded in 2007, having covered
11,663 deg$^2$. These data are part of the SDSS-I/II survey and are 
described in \citet{DR7} and references therein. SDSS-III added another
$\sim$3000 deg$^2$ of new imaging area in 2008. Pan-STARRS imaging
commenced in 2009 and continued through to 2013.  Hence, the addition
of the PS1 photometry to the SDSS photometry increases the baseline of
observations from $\approx$8 to $\approx$14 years.

    \subsubsection{SDSS}
    The SDSS \citep{York00} uses the imaging data gathered by a dedicated
    2.5m wide-field telescope \citep{Gunn06}, which collected light from a
    camera with 30 2k$\times$2k CCDs \citep{Gunn98} over five broad bands
    - {\it ugriz} \citep{Fukugita96} - in order to image 14,555 unique
    deg$^{2}$ of the sky. This area includes \hbox{7,500 deg$^{2}$} in the
    North Galactic Cap (NGC) and \hbox{3,100} deg$^{2}$ in the South
    Galactic Cap (SGC). The imaging data are taken on dark photometric
    nights of good seeing \citep{Hogg01} and are calibrated
    photometrically \citep{Smith02, Ivezic04, Tucker06, Padmanabhan08a},
    and astrometrically \citep{Pier03} before object parameters are
    measured \citep{Lupton01, Stoughton02}.
    The Eighth Data Release \citep[DR8;][]{DR8} provides updated photometric calibrations.

    The Stripe 82 region of SDSS (S82; 22h 24m $<$ R.A.\ $<$ 04h 08m
    and $|$Dec.$|<1.27$~deg) covers $\sim$300~deg$^2$ and has been observed
    $\sim$60 times on average to search for transient and variable objects \citep{DR7}.  These
    multi-epoch data have time scales ranging from 3 hours to 8 years and
    provide well-sampled 5-band light curves for an unprecedented number
    of quasars \citep[see][for examples of quasar variability studies
    based on S82 photometry]{ses07,schm10,ai10,mac10,meu11,but11}.
    
     In addition to the DR8 photometry provided on the SDSS website, 
     our analysis utilizes the S82 database of quasars in
     \citet{mac12}, which includes observations taken in nonphotometric
     conditions and recalibrated using the improved method of
     \citet{ive04}. While the latter dataset only includes point sources,
     we consider both resolved and unresolved observations from DR8, adopting PSF 
     magnitudes in each case\footnote{We adopt PSF magnitudes throughout our
       analysis, although ideally CMODEL (KRON) magnitudes should be used for
       extended sources in SDSS (PS1).  Since we are only interested in large
       magnitude changes in the central regions of any object, the
       differences in magnitude types should not appreciably affect our
       results. Also, we do not correct the magnitudes for Galactic
       absorption, as we are only interested in magnitude differences.}.
    We define a source to be in S82 if it is in the S82 database of 
     \citet{mac12} or in the (R.A., Dec.) range defined above.

        \subsubsection{Pan-STARRS1 3$\pi$ Survey}
        Our analysis includes imaging from the PS1 3$\pi$ survey
        \citep{kai02}, in particular the Processing Version 2 catalog
        available in a local DVO database (released January 2015).
        PS1 comprises a 1.8m telescope equipped with a 
        1.4-gigapixel camera.  Over the course of 3.5 years of the 3$\pi$ survey, up to four
        exposures per year in 5 bands, $g_{\rm P1}, r_{\rm P1}, i_{\rm P1},
        z_{\rm P1}, y_{\rm P1}$ have been taken across the full $\delta >
        -30^{\circ}$ sky \citep[for full details, see][]{Tonry12,met13}.  
        Each nightly observation consists of a pair of exposures 15~min apart 
        to search for moving objects. For each exposure, the PS1 3$\pi$ 
        survey has a typical 5$\sigma$ depth of 22.0 in the $g$-band \citep{ins13}.
        The overall 
        system, photometric system, and  the PS1 surveys  are described in
        \citet{Kaiser10}, \citet{Stubbs10}, and \citet{Magnier13}, respectively.

        \subsubsection{Catalina Sky Survey}
        While not included in our analysis, where instructive, we also show data from the
        Catalina Sky Survey second release \citep[CRTS;][]{dra09}. The CRTS magnitudes
        are based on unfiltered light but calibrated to a $V$-band zero-point. 
        We average the CRTS data in segments of 10 days for visual clarity, and apply 
        a constant offset $m_0$ so that the data match any simultaneous SDSS $g$-band observations.
    
    \subsection{Spectroscopic Data}
    We use the spectroscopic observations of quasars that are given in the
    SDSS Data Release Seven catalog (DR7Q) from \citet{sch10}.

    As described by \citet{ric02}, quasar target candidates are selected 
    for spectroscopic observations based on their optical colors and 
    magnitudes in the SDSS imaging data or their detection in the FIRST 
    radio survey \citep{Becker95}. Low-redshift, $z\lesssim 3$, quasar targets
    are selected based on their location in $ugri$-color space and the
    quasar candidates passing the $ugri$-color selection are selected to a
    flux limit of $i = 19.1$. High-redshift, $z\gtrsim 3$, objects are
    selected in $griz$-color space and are targeted to $i = 20.2$.
    Furthermore, if an unresolved, $i \leq 19.1$ SDSS object is matched to
    within 2'' of a source in the FIRST catalog, it is included in the
    quasar selection.
    
    The final quasar catalog from SDSS-I/II, based on the Seventh Data
    Release of SDSS \citep[DR7;][]{DR7}, is presented in \citet{sch10}.
    This catalog contains 105,783 spectroscopically confirmed quasars that
    have luminosities larger than $M_i = -22.0$.
    
    In order to look for significant changes in the BELs, we require
    (at least) a second epoch of spectroscopy. This is supplied by the
    Baryon Oscillation Spectroscopic Survey \citep[BOSS; ][]{Dawson13}
    which was part of the third incarnation of the SDSS
    \citep[SDSS-III;][]{Eisenstein11}.

    We apply no selection to the type of BOSS spectroscopic target
    that is utilised for the later epoch of spectroscopy; i.e., a DR7 SDSS
    quasar that is a ``changing-look'' candidate can be classified as a
    BOSS galaxy.  Indeed, of the final sample of 10 objects presented in this work, 
    only 3 are actually in the SDSS-III BOSS DR12 Quasar Catalog of
    {P{\^a}ris} et al.~(2016, in advanced prep.).  We give further details of the repeat spectroscopic
    targeting in the Appendix.

    The BOSS spectrographs and their SDSS predecessors are described in
    detail by \citet{Smee13}. In brief, there are two double-armed
    spectrographs that are significantly upgraded from those used by
    SDSS-I/II. Exposed to a minimum signal-to-noise ratio of $\sqrt{10}$ in $g$ and $\sqrt{20}$ in $i$  
    \citep[$\sim$1.6~hr/plate;][]{Dawson13}, they cover the wavelength range $3600\,$\AA\ to
    $10,400\,$\AA\ with a resolving power of 1500 to 2600 \citep{Smee13}.
    In addition, the throughputs have been increased with new CCDs,
    gratings, and improved optical elements, and the 640-fibre cartridges
    with $3"$ apertures have been replaced with 1000-fibre cartridges with
    $2"$ apertures. Ultimately, the throughput of the BOSS spectrographs
    are considerably greater (in the red and the blue), and span a greater
    wavelength range, than the original SDSS instruments. 

    The BOSS spectra presented in this work all have
    \small{LAMBDA\_EFF}\normalsize$ = $5400\AA\ \citep{Dawson13}, i.e.,
    the SDSS plate holes were drilled to maximize the signal-to-noise at
    5400\AA, and therefore do not need the spectrophotometric corrections
    from \citet{mar15}.  Also, the BOSS spectra presented here do not have
    the \small{PROGRAM$ = $APBIAS} \normalsize target flag, which would
    indicate an offset in the fiber position with respect to the earlier
    SDSS spectrum.

    \subsection{Multi-wavelength Coverage} 
    We cross-matched our superset of quasars with various radio and
    X-ray catalogs. We use the combined radio catalog of
    \citet{Kimball_Ivezic14}, which includes sources from five radio
    catalogs (FIRST, NVSS, GB6, WENSS, and VLSSr), to help identify blazar
    contaminants in S82 during the selection process. We check the latest
    release of the XMM-{\it Newton} serendipitous source catalog
    \citep{ros15} and the {\it Chandra} Source Catalog
    \citep[CSC;][]{eva10} for archival X-ray observations of any interesting
    objects from our search.

\section{Sample Selection}\label{sample}

Our superset is any object
listed in the DR7Q catalog of spectroscopically confirmed quasars,
which includes both point-sources and resolved objects with $M_i<-22$.
To select quasars that may have varying spectral features, we
quantify the photometric properties of this spectroscopic quasar
dataset and assume that significant BEL changes will be associated
with a significant change in flux.  We use the $g$-band SDSS
photometry and extend the time baseline from 10 to 15 years by including 
$g$-band PS1 photometry in our analysis\footnote{The SDSS $g$ filter is 
close enough to the $g_{\rm P1}$ filter in overall response that we can 
ignore any color terms.}. Since 
our aim is to find changing-look quasars, we search for quasars that, 
along with the earlier spectrum in SDSS DR7, have a later spectrum in BOSS.

Initially, we limit our sample to the S82 region, so that a
well-sampled light curve exists for each object, making it easier to
identify true large-amplitude photometric variability. 
There are 9474 quasars in S82, including extended sources
which are not in the point source catalog of \citet{mac12}.  Motivated
by the light curve for J0159+0033, we search for quasars that show at
least a 1.0~mag dimming or brightening in the $g$-band among any
observations in the combined SDSS and PS1 light curve\footnote{For a similar search but for large-amplitude 
(1.5~mag) nuclear brightening in resolved SDSS galaxies, see \citet{law12b}, 
which utilizes results from the PS1 Faint Galaxy Supernova 
Survey available at {\tt http://star.pst.qub.ac.uk/sne/ps1fgss/psdb/}.}.
For objects with at least ten photometric data points, light curve 
outliers are flagged as being 0.5 mag away in $g$ from the light curve 
running median ($\sim$30\% of the sample). Since our aim is to find 
large, gradual changes in flux without a significant amount of 
contamination due to poor photometry, we reject these light curve 
outliers during the variability selection.  This selects 
1692 objects with $|\Delta g|>1$~mag and photometric
uncertainties $\sigma_g<0.15$~mag. Approximately 15\% of these were
observed again with the BOSS spectrograph; we focus on these 287
objects. Thirty-six objects in this subsample are detected in the radio, 
and of these, three were clearly blazars, as they were radio sources and 
exhibited fast and large-amplitude variability \citep[2--3 mag within months;
e.g.][]{rua12}.  We do not consider these three objects in our further
analysis, as we are interested in BEL changes unrelated
to blazar activity.  After visually searching through all SDSS/BOSS
spectra for BELs that are clearly present in one epoch but not
another, we identify seven quasars from S82 in which at least some
BELs satisfy this criterion.

\begin{figure}
\centering
\includegraphics[scale=.53,trim=1.5cm 0cm 0cm 0cm,clip=true]{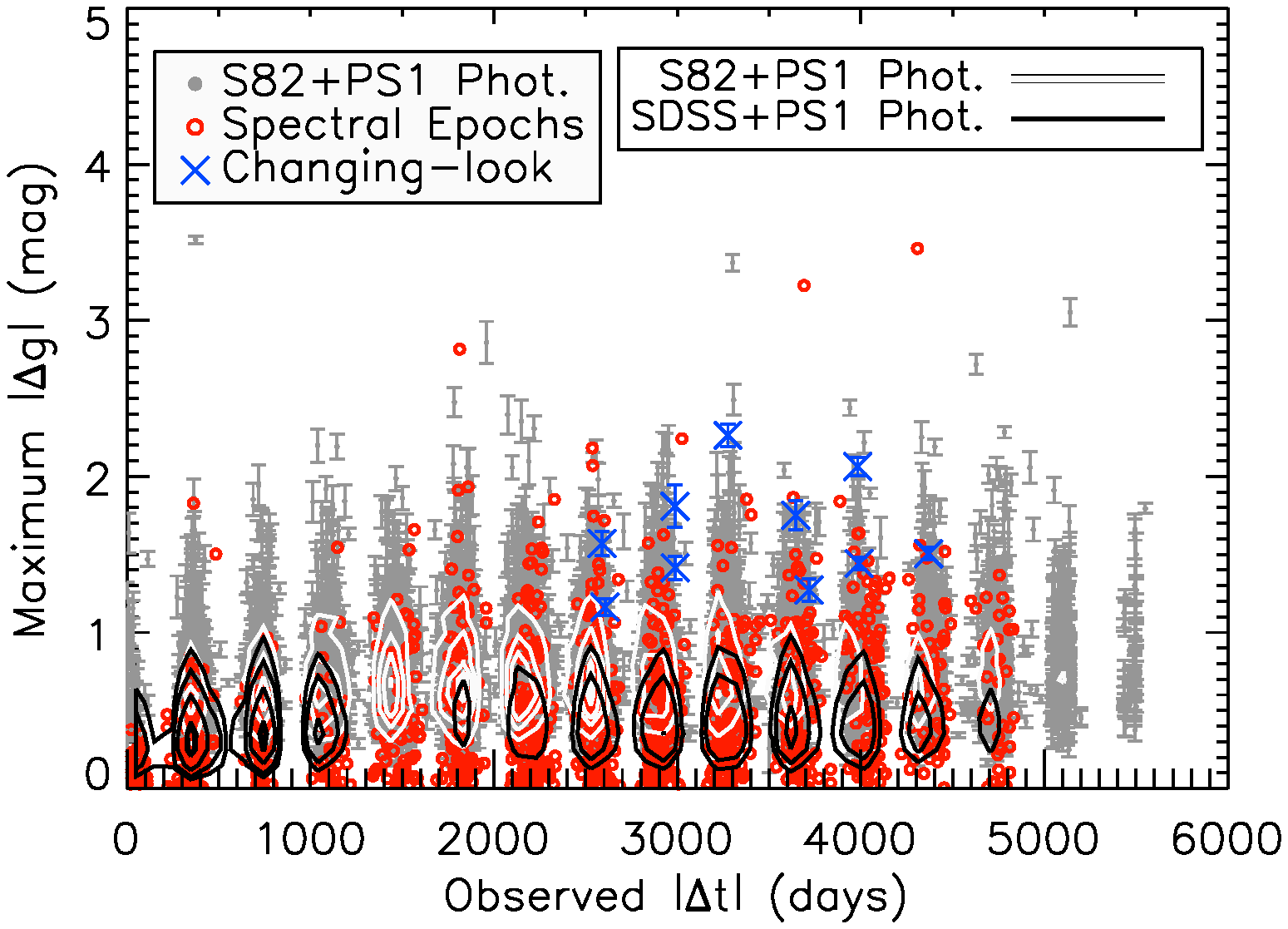}
\includegraphics[scale=.53,trim=1.5cm 0cm 0cm 0cm,clip=true]{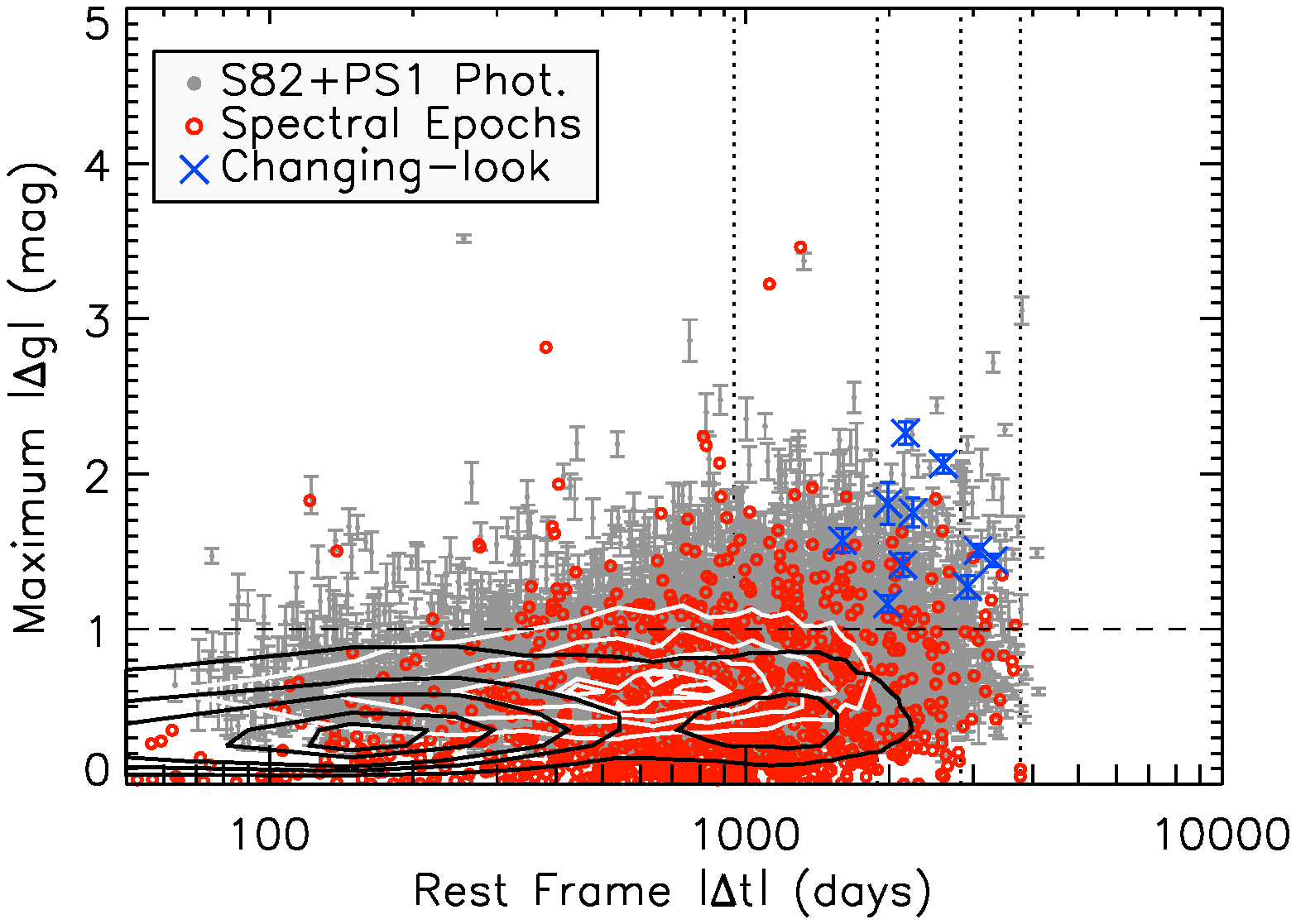}  
\caption{Distribution of the maximum magnitude difference $|\Delta g|_{\rm max}$ versus time lag $|\Delta t|$.
The black contours show the distribution for the superset of quasars (DR7Q), which is based on SDSS and PS1 photometry.  
 The white contours are the same but for the subsample of quasars in S82, which are shown as grey data points (one point per S82 quasar). 
The contours show regions containing 5, 10, 25, 50, and 75\% of the data.
The subsample of DR7Q with repeat spectra and $|\Delta g|>1$ photometric changes are shown by open red circles, but now showing the $|\Delta g|_{\rm max}$ spanned by the \emph{spectroscopic} epochs versus the corresponding time between spectra. For objects with multiple spectra, we choose the two epochs spanning the largest $|\Delta g|$ for display here.
While the data are clumped into ``seasons'' in the top panel, the distribution is smoothed out when switching to rest-frame time lag in the bottom panel.
Our final selection is limited to quasars with repeat spectra with $|\Delta g|_{\rm max}>1$ (indicated by the horizontal dashed line). 
The final sample of 10 objects are plotted as blue crosses. 
}
\label{fig:dt_dg}
\end{figure}

We then extend our search to the entire SDSS footprint, which contains
105783 quasars in DR7Q. Of the 105746 quasars ($>99$\%) which have PS1
detections, 6348 have shown at least a 1.0~mag change in
their $g$-band light curves. Of these, 1011 have BOSS spectra, 
which includes the 287 quasars from S82. After visually inspecting
each spectrum, we find three additional quasars with disappearing
BELs. The final yield is higher for the S82 sample due to the
improved cadence; we are able to more efficiently identify high-amplitude variability
as well as more reliably identify spurious data points\footnote{\footnotesize 
Without having a well-sampled light curve,
outliers due to poor photometry are more difficult to identify by our
algorithm and therefore can cause the object to pass the $|\Delta
g|>1$~mag criterion.}. The distributions of the time scales and
magnitude changes involved are shown in Fig.~\ref{fig:dt_dg}. This
selection algorithm skews our search to those objects showing BEL
changes over roughly ten years, since this is the timespan between
SDSS and BOSS spectra, although the rest-frame timescales probe down to 
shorter timescales (bottom panel). The improved time coverage of the S82 regions
can also be seen from the contours in the panels; the S82 sample (in
white) fills in gaps in $|\Delta t|$ while reaching to larger $|\Delta
g|$.

Our sample selection is given in Table~\ref{tab:selection}.

\begin{table}
  \begin{center}
    \begin{tabular}{lrr}
      \hline
      \hline
      Selection     & Total \# & In S82\\
      \hline 
      SDSS Quasars in DR7Q                       & 105783 & 9474 \\
      \quad  with BOSS spectra   &  25484 & 2304  \\
      \qquad and $|\Delta g|>1$~mag and $\sigma_g<0.15$~mag & 1011 & 287 \\
      \quad \qquad and that show variable BELs &  10 & 7 \\
      \hline
      \hline
    \end{tabular}
    \caption{Selection of spectroscopically variable quasars.}
    \label{tab:selection}
  \end{center}
\end{table}

\section{The Changing-Look Quasars}\label{results}

Our initial search through the S82 quasars yielded the following
results: (i) significant BEL changes are seen on long timescales
($\sim$2000 to 3000~days in the rest frame) in the selected sample;
(ii) these changes are associated with large ($|\Delta g|\sim 1$)
amplitude changes in the photometry, and (iii) emerging (disappearing)
BEL features correspond to continuum brightening (dimming). Given the
extra temporal information provided by S82 light curves, our selection
algorithm could more easily identify large-amplitude outbursts and
reliably reject spurious data points, yielding seven objects of
interest in S82.

When extended to the full SDSS footprint, where the inclusion of the
PS1 3$\pi$ photometry generated lightcurves for $>$99\% of the DR7Q
quasars, our combined search yielded three additional objects
(Table~\ref{tab:selection}). We present all 10 objects here: four that
show {\it appearing} BEL features, five that show {\it
disappearing} BELs\footnote{One object, SDSS J214822.25+011217.6, is 
not considered part of our sample although it exhibited a disappearing 
BEL; the disappearance was due to the appearance of a broad absorption 
line, which is a different phenomenon than the changing-look 
behavior studied here \citep[see][]{FilizAk12}.}, and one that shows 
evidence for both.

\begin{figure}
\centering
\includegraphics[scale=.55,trim=1.5cm 0cm 0cm 0cm,clip=true]{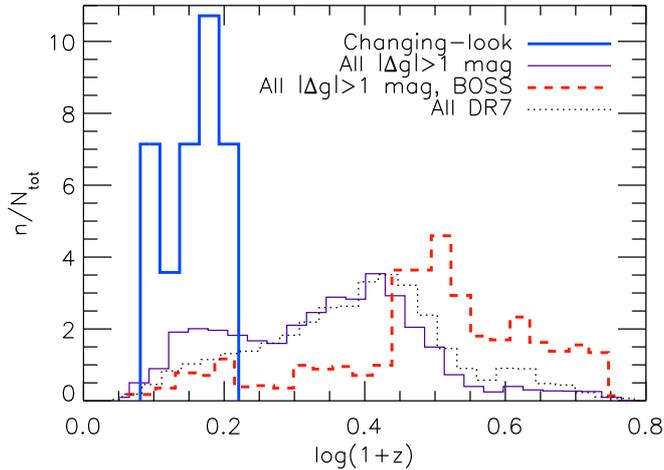}
\caption{Normalized redshift distributions for the sample of changing-look quasars (in blue), the full sample of  $N_{\rm tot}=6348$ highly variable quasars (purple), the subset with BOSS spectra (dashed red;  $N_{\rm tot}=1011$), and the entire DR7 quasar catalog (dotted; $N_{\rm tot}=105,783$). $n$ indicates the number of points in a
bin divided by the bin width, and each histogram has unit area. }
\label{fig:zdist}
\end{figure}

\begin{figure}
\centering
\includegraphics[scale=.55,trim=1.5cm 0cm 0cm 0cm,clip=true]{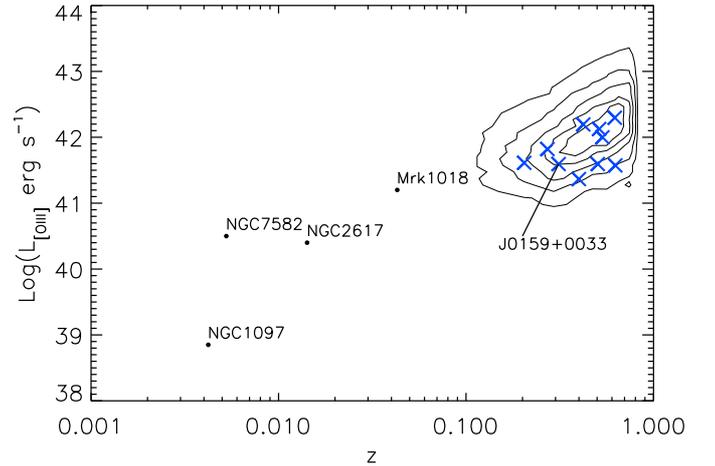}
\caption{\oiii luminosity versus redshift for the DR7Q sample (contours), our final sample of 10 (blue crosses), and examples of previously known changing-look AGN \citep[black points; cf. Fig.~1 in][]{lam15}. For the DR7Q sample, we adopt values from \citet{she11}. }
\label{fig:zldist}
\end{figure}

Our final sample of changing-look quasars is listed in
Table~\ref{tab:objs}, and the redshift distribution is compared to the
full quasar sample in Fig.~\ref{fig:zdist}.  We note that all of our
objects are at $z<0.63$, but this is potentially a selection effect,
as we discuss in Section~\ref{disc}. However, this sample extends the 
range of known changing-look AGN to $z=0.63$ at quasar luminosities. 
In Fig.~\ref{fig:zldist}, the 5007\AA\ \oiii luminosity is shown 
as a function of redshift for the full DR7Q sample and our final objects.
Examples of previously studied changing-look AGN are also shown for 
comparison. The \oiii luminosity is often used as a proxy of the intrinsic 
AGN luminosity \citep[e.g.,][]{kau03,hec04}, so that Fig.~\ref{fig:zldist} 
is a comparison of the intrinsic brightness and redshift of changing-look AGN. 

In the following sections, we compute the flux deviation between two
spectra at any given wavelength,
$N_{\sigma}(\lambda)=(f_2-f_1)/\sqrt{\sigma_2^2+\sigma_1^2}$
\citep[e.g.,][]{FilizAk12}, to determine the significance (in units of
$\sigma$ per spectral pixel) of a BEL change. In particular, we assess 
the significance of a BEL change by comparing its flux deviation to that of the
underlying continuum at that wavelength. Note that the significance of BEL changes 
will be higher than that quoted here when $N_{\sigma}(\lambda)$ is 
integrated over the pixels spanning the BEL. 
In the few cases where there are more than two spectra available, we adopt the two 
spectra with the largest time lag unless otherwise stated. 

In each case, the difference spectrum $|\Delta f_{\lambda}|=|f_{\rm BOSS}-f_{\rm SDSS}|$ 
is presented.  $|\Delta f_{\lambda}|$ is fit as a power-law after masking out the 
H$\alpha$, H$\beta$,  H$\delta$, H$\gamma$, \mgii and \small{HeI} \normalsize BELs 
and allowing the normalization to vary. The best-fit power-law indices 
$\beta$ are listed in Table~\ref{tab:objs} and are based on the 
spectra listed in the preceding columns. We compare the best fits to a $f_{\nu}\propto \nu^{1/3}$ 
power law since this form is expected if the variable component resembles a standard thin disk \citep{SS73}.
 

\begin{table*}
  \caption{SDSS Quasars with emerging or disappearing BELs. 
    $\Delta g$ is the largest magnitude change observed, and $\Delta t_{\rm RF}$ is
    the time span for this change in the rest frame of the quasar.  J015957.64$+$003310.4 is the (first)
    changing-look quasar discovered by \citet{lam15}. Max($\Delta g$) is
    defined such that positive values imply a dimming in the light curve.
    The last column lists the best-fit power-law index $\beta$ for the difference spectrum.}
  \label{tab:objs}
  \begin{tabular}{l ccc c ccc}
    \hline
    \hline
    Name  &  $z$ & Max($\Delta g$) & $\Delta t_{\rm RF}$ & BEL & $({\rm MJD|plate|fiber})_{\rm 1}$ & $({\rm MJD|plate|fiber})_{\rm 2}$ & $|\Delta f_\nu| \propto \nu^\beta$\\
    (SDSS J) &    &  & (days)  &  behavior &   &  &   \\
    \hline
    002311.06+003517.5 & 0.422 &$-1.50 \pm  0.04$&3072  & Appear   & $51816|0390|0564$ & $55480|4219|0852$ & $ \phantom{-}0.04  \pm    0.02  $\\
    015957.64+003310.4 & 0.312 &$ 1.16 \pm  0.06$&1985  & Disappear& $51871|0403|0549$ & $55201|3609|0524$ & $ \phantom{-}0.27  \pm    0.02  $\\
    022556.07+003026.7 & 0.504 &$ 1.81 \pm  0.14$&1985  & Both     & $52944|1508|0556$ & $55445|3615|0617$ & $ \phantom{-}0.16  \pm    0.03  $\\
    022652.24-003916.5 & 0.625 &$ 1.75 \pm  0.09$&2242  & Disappear& $52641|1071|0281$ & $56577|6780|0339$ & $ \phantom{-}0.2   \pm    0.1   $\\
    100220.17+450927.3 & 0.400 &$ 1.41 \pm  0.07$&2134  & Disappear& $52376|0943|0310$ & $56683|7284|0122$ & $-0.20  \pm    0.02  $\\
    102152.34+464515.6 & 0.204 &$ 1.44 \pm  0.04$&3313  & Disappear& $52614|0944|0603$ & $56769|7386|0410$ & $ \phantom{-}0.175 \pm    0.007 $\\
    132457.29+480241.2 & 0.272 &$ 1.27 \pm  0.07$&2923  & Disappear& $52759|1282|0045$ & $56805|7406|0527$ & $ \phantom{-}0.86  \pm    0.02  $\\
    214613.31+000930.8 & 0.621 &$-1.57 \pm  0.08$&1597  & Appear   & $52968|1107|0358$ & $55478|4196|0774$ & $ \phantom{-}0.1   \pm    0.1   $\\
    225240.37+010958.7 & 0.534 &$-2.06 \pm  0.06$&2596  & Appear   & $52174|0676|0442$ & $55500|4294|0045$ & $-0.45  \pm    0.08  $\\
    233317.38-002303.4 & 0.513 &$-2.26 \pm  0.07$&2164  & Appear   & $52199|0681|0114$ & $55447|4212|0312$ & $ \phantom{-}0.75  \pm    0.07  $\\
    \hline
    \hline
  \end{tabular}
\end{table*}

    \subsection{Appearing BELs}
\begin{figure*}
\centerline{
\includegraphics[scale=.45]{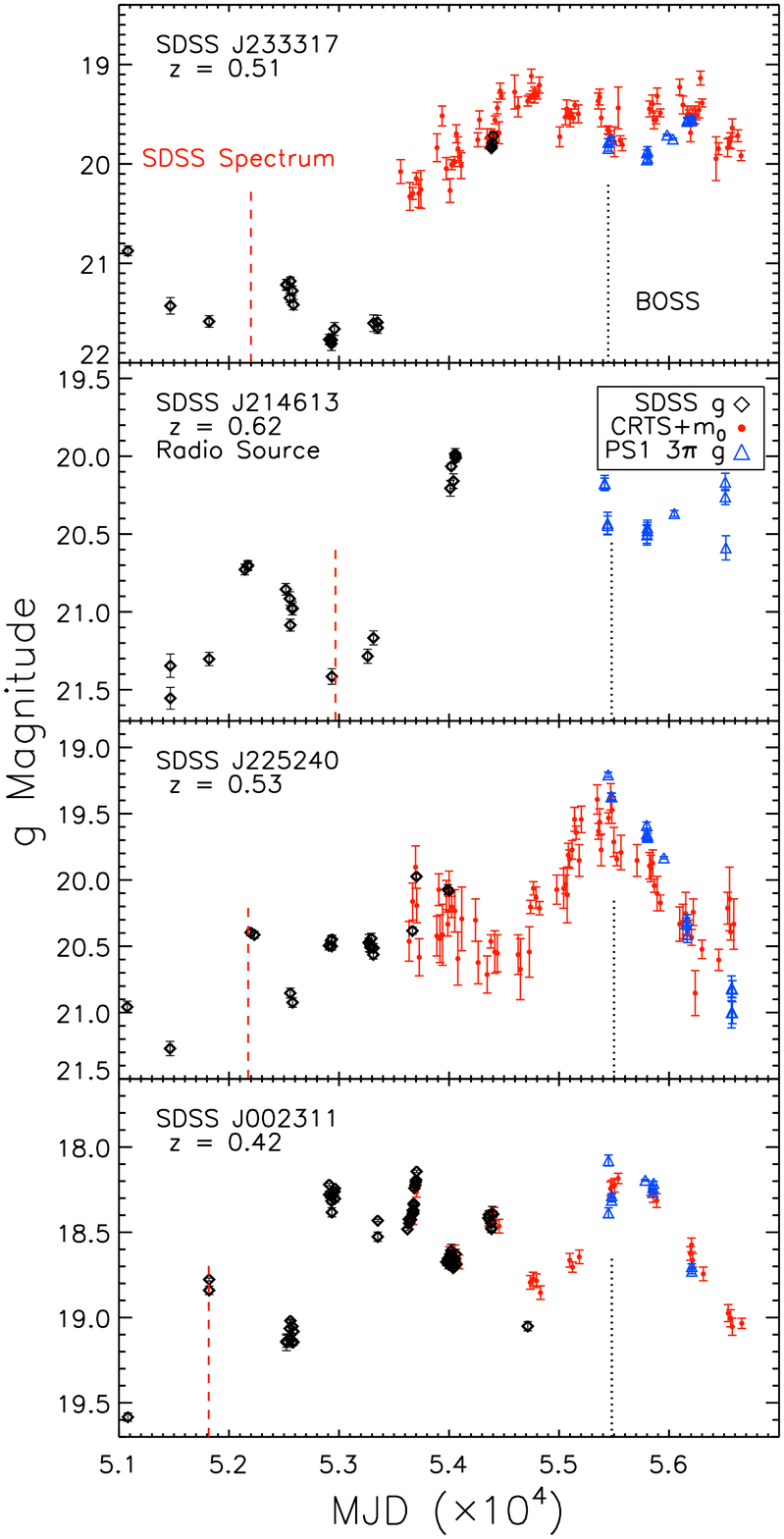}
\includegraphics[scale=.45]{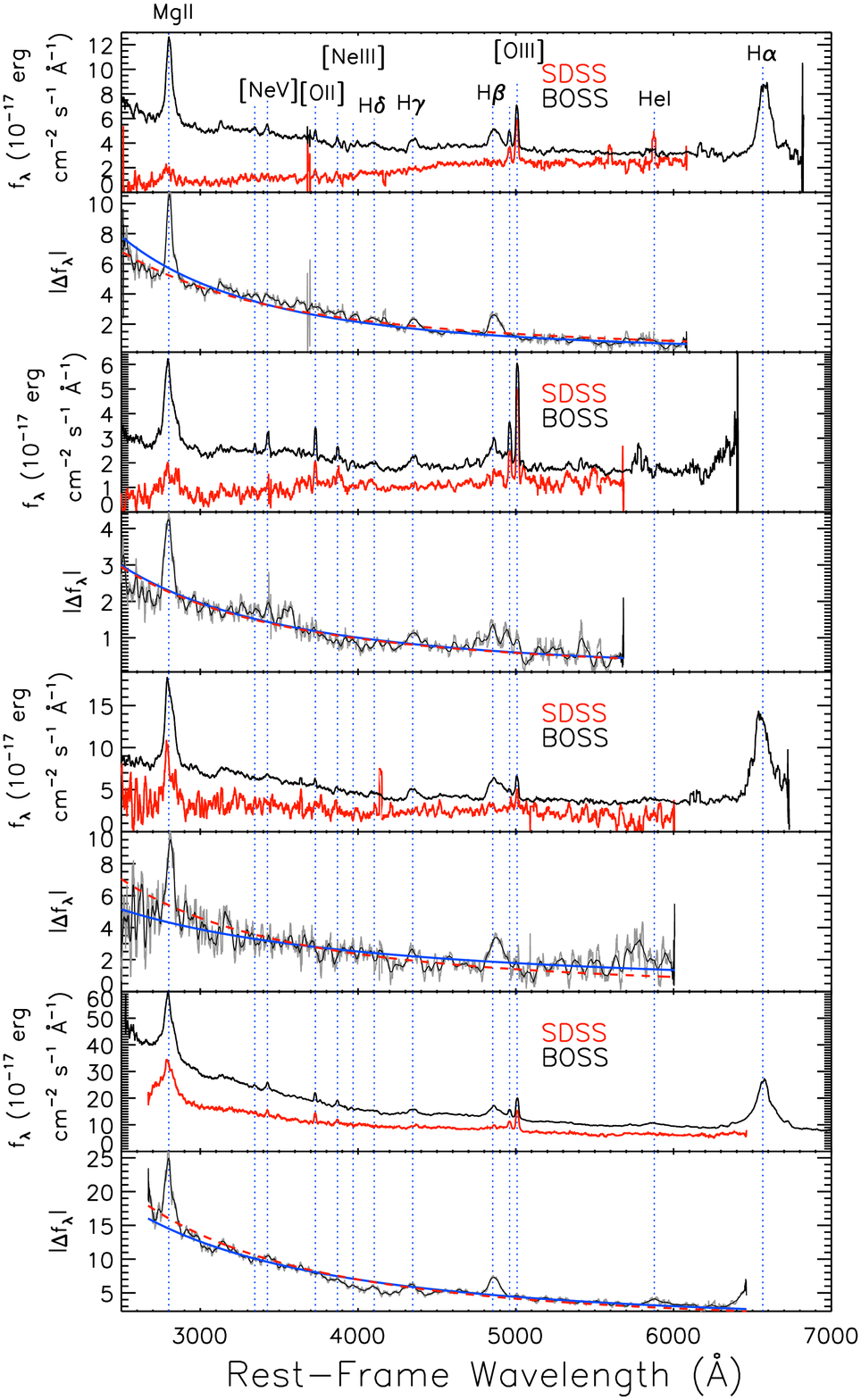}  }

\caption{ Quasars that show evidence for emerging BELs.  The light curves in the left panel show the SDSS $g$-band (black diamonds), PS1 $g$-band (blue triangles), and CRTS photometry  (red dots).  The right panels show the SDSS and BOSS spectra in red and black, respectively, with epochs indicated by the (color-coded) vertical dotted lines in the light curve. The flux difference  $|\Delta f_{\lambda}|$ between the BOSS and SDSS spectra is shown in the right lower subpanels, where the blue curve is the best-fit power-law $f_{\nu}\propto \nu^{\beta}$, and the red-dashed curve is a power-law with $\beta=1/3$ which is expected for a standard thin disk.   } 
\label{fig:appear}
\end{figure*}

    Four objects that show evidence of appearing BELs are plotted in
    Fig.~\ref{fig:appear}. In all cases, the flux
    increased dramatically ($|\Delta g|> 1$~mag). The light curves
    in the top two panels of Fig.~\ref{fig:appear} show a ``flat-topped''
    behavior, i.e., rising over $\sim$1000 days in the observed frame to a 
    constant luminosity. J214613.31+000930.8 (second panel) is the
    only object from our final sample that has a radio detection.
    In all cases, H$\beta$ is absent from the first spectrum while \mgii is
    observed at low signal-to-noise. In general, however, the significance for
    appearing BELs in our sample is not very high ($\lesssim 3\sigma$ per spectral pixel for 
    all H$\beta$ transitions) for two reasons. First, in
    order to be included in our search, the source must be a BEL quasar in
    the initial DR7 spectrum, thus making any further BEL brightening less
    significant. Second, the improved BOSS spectrograph provides higher
    quality spectra than the SDSS spectrograph, so
    if the source is faint in the earlier spectroscopic epoch, its
    spectrum will be correspondingly relatively noisy. 
    However, with these caveats in mind, it is notable that our search 
    produced a similar number of appearing and disappearing BEL cases.

    \subsection{Disappearing BELs}
\begin{figure*}
\centerline{
\includegraphics[scale=.45]{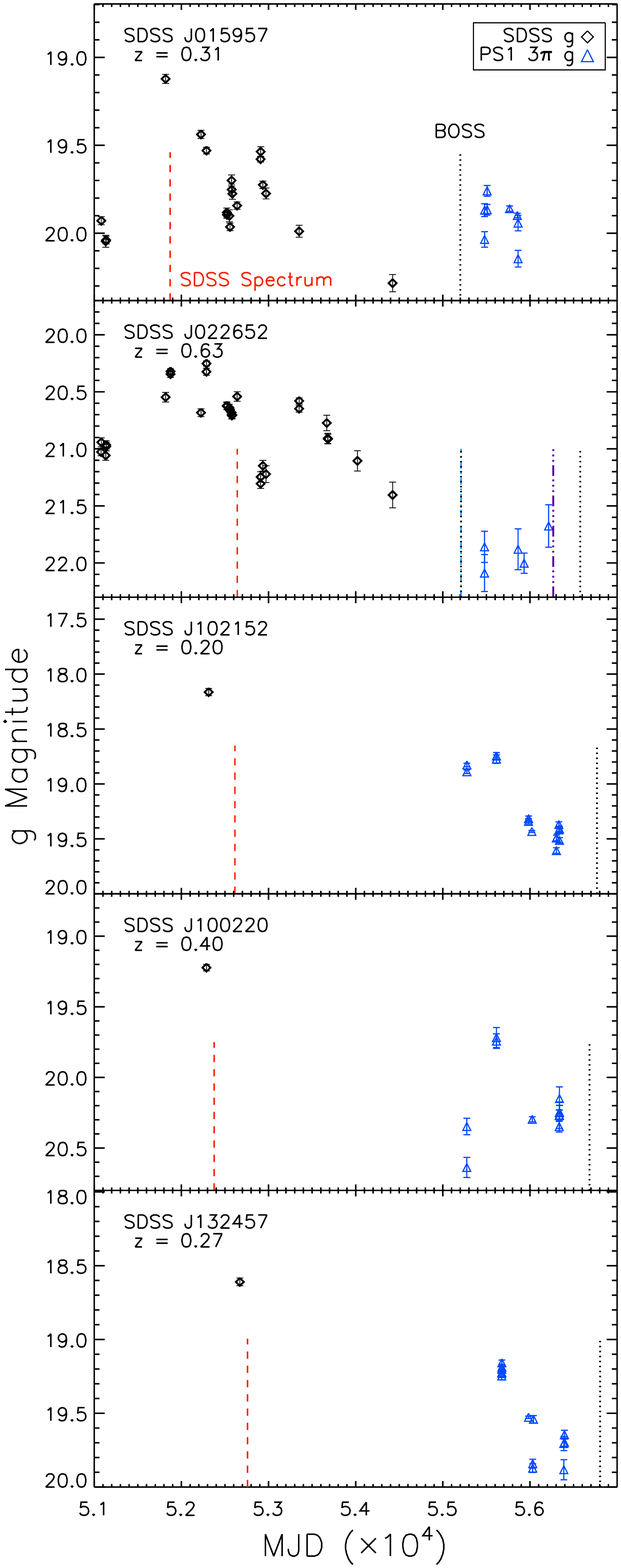}
\includegraphics[scale=.45]{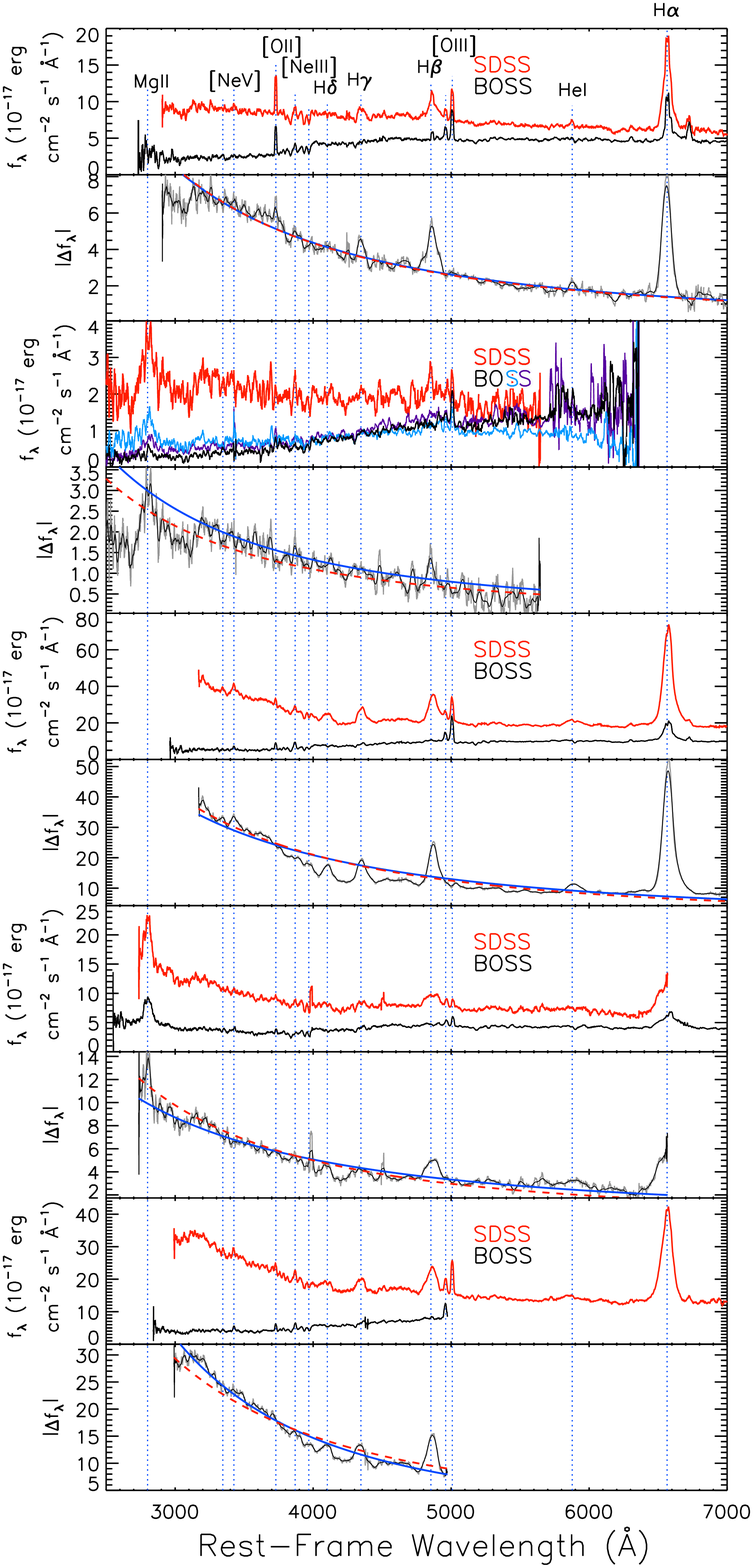}  }

\caption{ Quasars with disappearing BELs. The top panels show J0159+0033, which was discovered in \citet{lam15}.  The cyan and purple spectra in the second (right) panel show additional BOSS epochs at MJD$=55209$ and 56267, respectively. }
\label{fig:disappear}
\end{figure*}

    The five objects that show evidence of disappearing BELs are plotted in
    Fig.~\ref{fig:disappear}. In all cases, the $g$-band flux dropped significantly
    from the SDSS spectroscopic epoch to the BOSS epoch.

    We recover J0159+0033 in our search (top panel of Fig.~\ref{fig:disappear}), which
    shows a vanishing H$\beta$ at $4\sigma$ significance.  Among our 10 final objects,  
    this is the only one present in the XMM or {\it Chandra} catalog.  
    J022652.24$-$003916.5, shown in the second panel, showed a similar behavior, and 
    multiple BOSS spectra reveal the object in an intermediate phase between 
    MJD$\approx$55200 and 56250 (cyan and purple spectra). In this case, the \mgii BEL is 
    barely present.  However, the BEL disappearance in this object is at low significance 
    ($<2\sigma$) since the source is faint ($20.5<g<22$). The remaining panels show three 
    objects from outside S82 where H$\beta$ vanishes at $>3\sigma$ significance.  The object 
    J102152.34+464515.6 (hereafter J1021+4645), shown in the third panel, demonstrates a 
    highly significant ($8\sigma$) change from a Type 1.0 to a Type 1.9 AGN, and we 
    elaborate on this object in Section~\ref{disc}. In the dim-state spectrum for 
    SDSS~J100220.17+450927.3, the broad \mgii line is still present, but the H$\beta$ 
    line is absent.

    The BOSS spectrum for J132457.29+480241.2, shown in the last panel, was 
    unrecoverable redward of H$\beta$ due to data extraction issues associated with that 
    particular BOSS fiber. However, an additional spectrum was obtained in January 2015 (${\rm MJD} = 57036$) by \citet{rua15} using the 3.5m telescope at the Apache Point Observatory (APO), and it is presented in the Appendix of that paper.  The H$\beta$ BEL is present in the APO spectrum but diminished with respect to the SDSS spectrum.  However, assuming that the measured BOSS flux is accurate, the APO spectrum was obtained when the object had rebrightened. 

    \subsection{Both Appearing and Disappearing BELs} \label{both}
\begin{figure*}
\centerline{
\includegraphics[scale=.4]{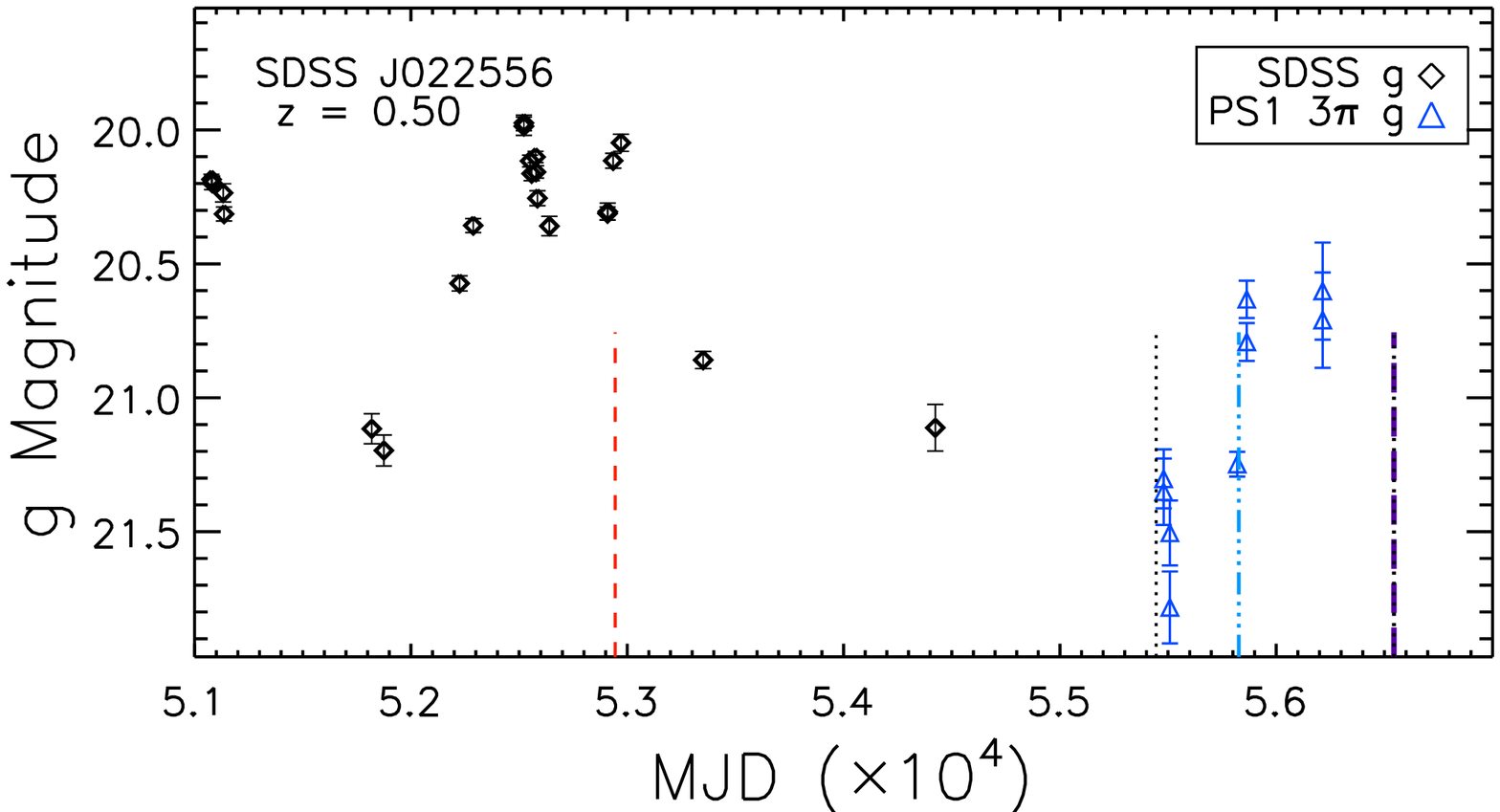}
\includegraphics[scale=.4]{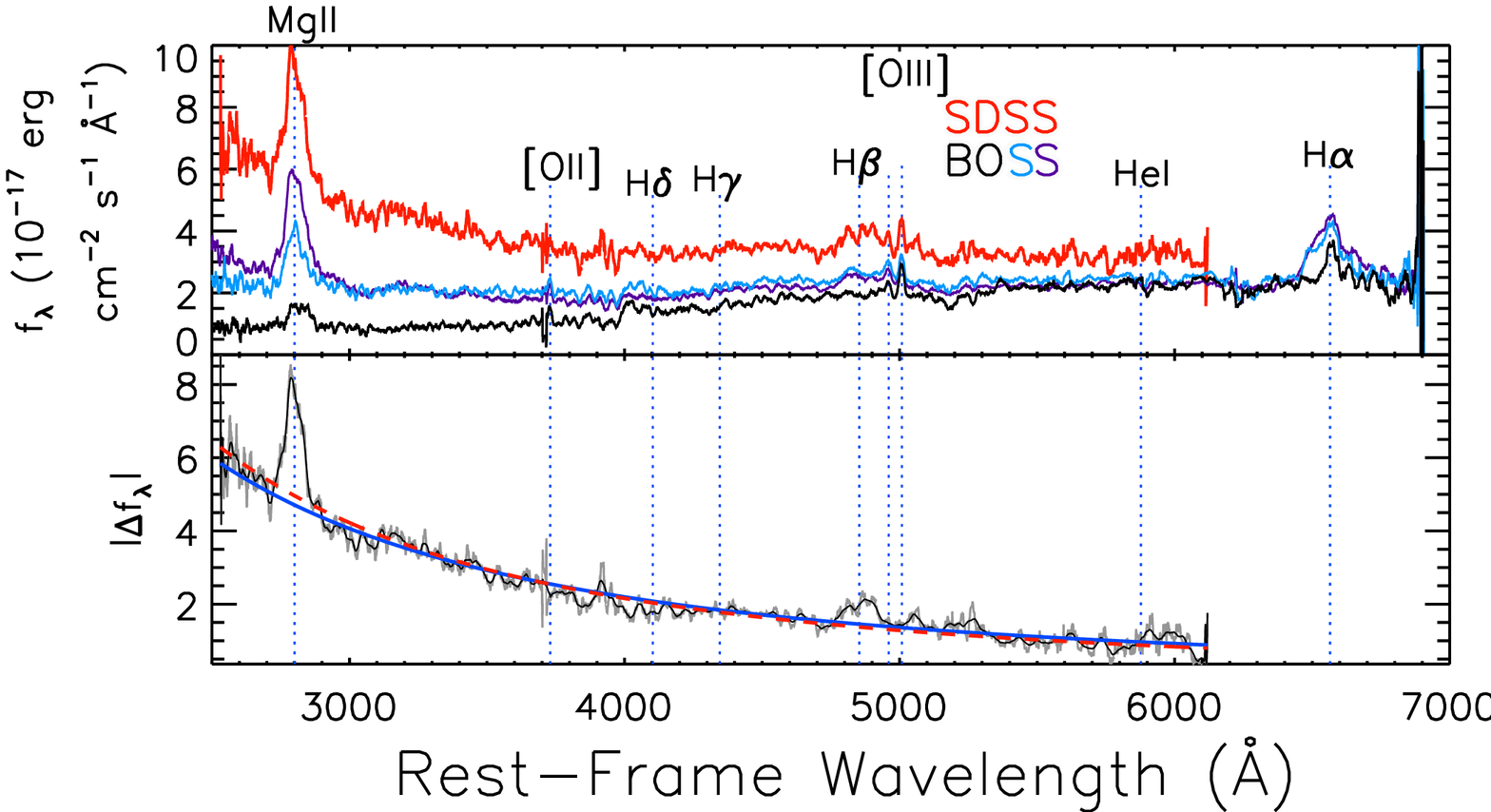}  }
\centerline{
\includegraphics[scale=.4]{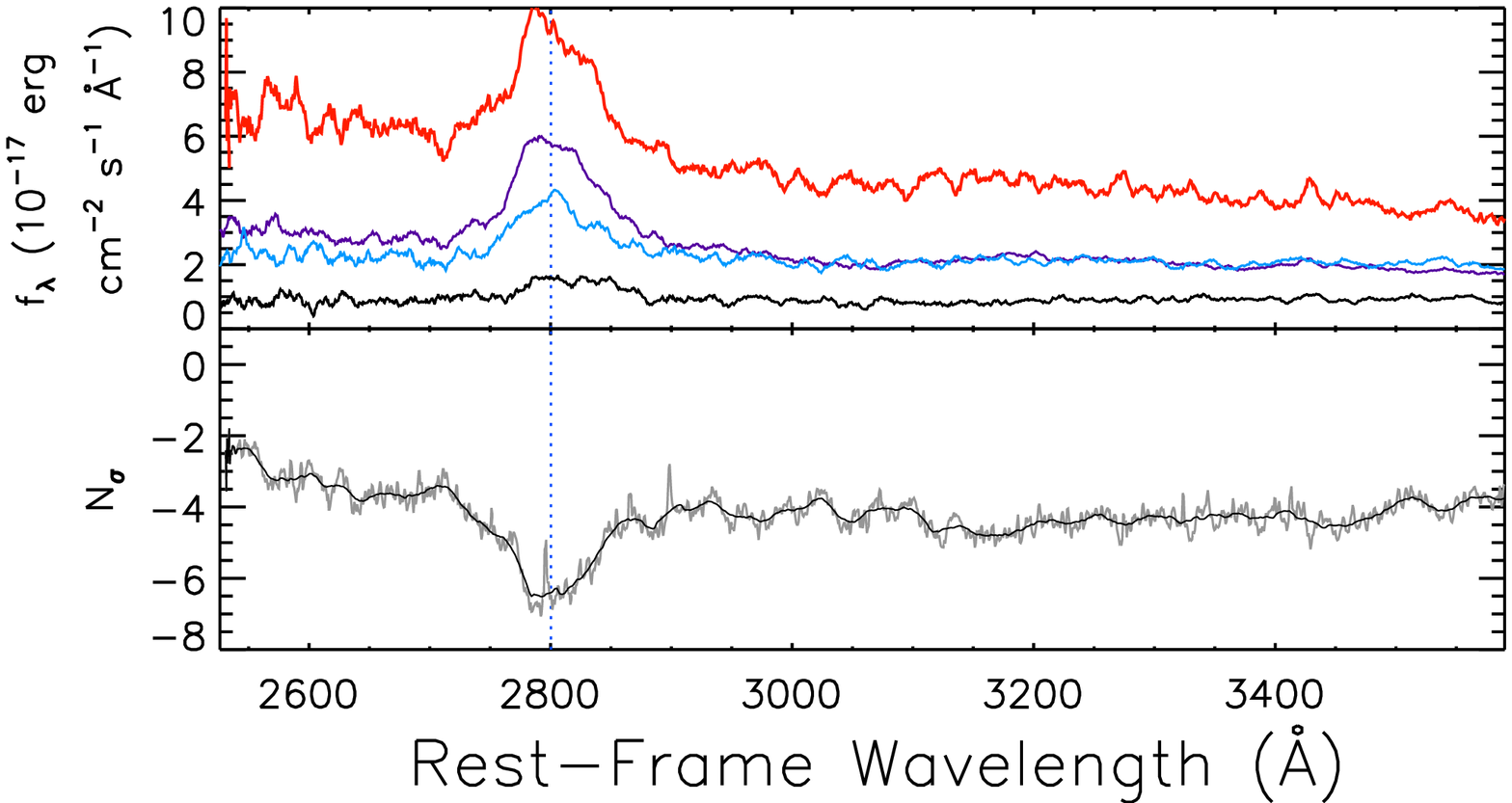}  
\includegraphics[scale=.4]{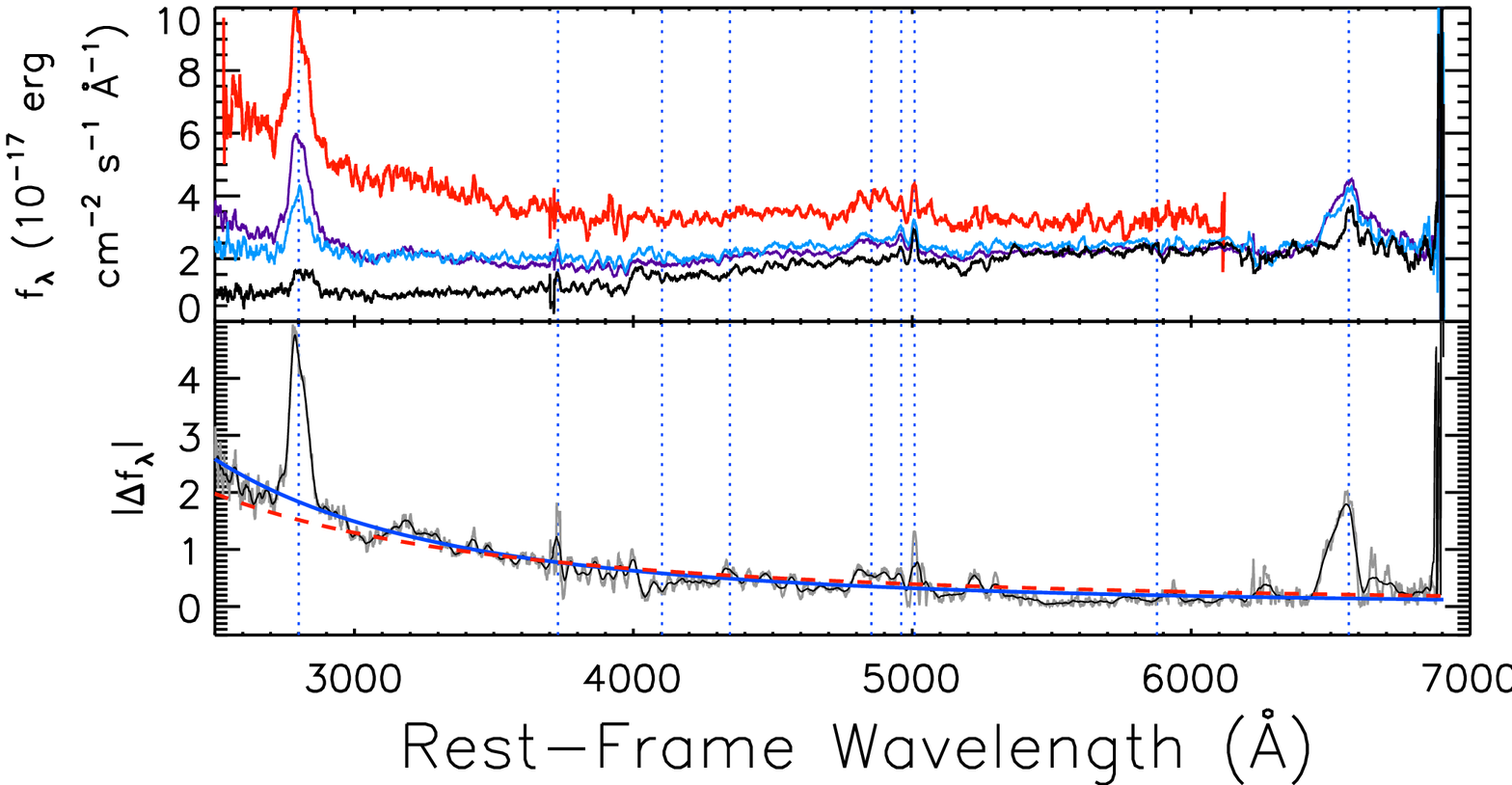}  }
\caption{A quasar exhibiting disappearing and reappearing BELs.  In the lower left panel, a zoom-in on the \mgii line shows profile changes, and the flux deviation $N_{\sigma}=(f_2-f_1)/\sqrt{\sigma_2^2+\sigma_1^2}$ between the early BOSS (black, $f_2$) and SDSS (red, $f_1$) spectra is shown in the bottom panel.  The lower right panel is identical to the upper right panel except that the flux difference $f_3 - f_2$ between the latest BOSS (purple, $f_3$) and earlier BOSS (black, $f_2$) spectra is shown.  The epochs for the red, black, cyan, and purple spectra are MJD$=52944$, 55445, 55827, and 56544, respectively. }
\label{fig:both}
\end{figure*}

    One object, SDSS~J022556.07+003026.7 (hereafter J0225+0030), shows a significant 
    evolution of the BELs as the source dims and rebrightens (see 
    Fig.~\ref{fig:both}). The 
    multiple BOSS spectra\footnote{J0225+0030 was observed multiple times due to
      it being on the extra deep BOSS plates 3615 and 3647, see {\tt http://www.sdss.org/dr12/spectro/special\_plates/}. There were 12 additional epochs of spectroscopy which are not shown in Fig.~\ref{fig:both}.}
    reveal a complete disappearance and re-emergence of H$\beta$, although
    this is at low significance due to low signal-to-noise.  
    \mgii shows a similar behavior, but it retains a broad component throughout.
    While outside the range of the SDSS spectrum, 
    in the BOSS spectra the broad H$\alpha$ component shows a clear 
    increase over the course of just one
    year, going from MJD$=$55445 (in black) to 55827 (in cyan).  Similar
    rapid BEL changes have also been observed in Seyfert galaxies such as NGC~7603 
    \citep{toh76} and Mkn~110 \citep{bis99}. 
    In the \mgii BEL profile shown in the lower left panel, the red wing diminishes faster
    than the blue wing which is broadly consistent with infalling gas. Such profile
    changes may provide a tool to study the structure of the BLR and
    warrants a future detailed study.

    Here, we are interested in the response of the  \mgii line to the underlying NUV 
    continuum in comparison to what is typically observed, since atypical behavior 
    may indicate a rare, more extreme event.  
    We measure the EW of \mgii and find it to be constant 
    ($\approx 100$\AA) over the large drop in local continuum flux going from 
    $5.2\times10^{-17}$~erg~cm$^{-2}$~s$^{-1}$~\AA$^{-1}$ in MJD$=$52944 to 
    $0.9\times10^{-17}$~erg~cm$^{-2}$~s$^{-1}$~\AA$^{-1}$ in 55445.
    For the few reverberation mapped sources with \mgii coverage, the \mgii line is typically unresponsive 
    \citep[e.g.,][]{cla91,kro91,goa93,cac15}, meaning that the line flux is constant as the 
    continuum level varies. The result is an intrinsic Baldwin effect 
    where the EW decreases with increasing continuum flux in a given AGN. 
    Since the \mgii EW remains constant 
    \citep[leading to a negligible Baldwin effect, as seen for an ensemble of objects in][]{die02}, it seems 
    that the \mgii BEL response in J0225+0030 is more pronounced than in typical AGN. 

    We elaborate on J0225+0030 in Section~\ref{0225}.

\subsection{Continuum Changes}

While the majority of highly-variable quasars in our search did not exhibit emerging or disappearing BELs, they often showed clear changes in the continuum slope in a bluer-when-brighter fashion, as is expected for quasars\footnote{We do not quantify the color changes for the overall sample of 1011 objects, as this is outside the scope of the present work.  For such a study, we refer readers to \citet{rua14}.} \citep[e.g.,][]{sch12}.  
In the 10 cases presented here, clear continuum slope changes were observed along with the BEL changes based on the ratio between spectra.

The difference spectra, shown in the lower right subpanels of Fig.s~\ref{fig:appear}--\ref{fig:both},  
eliminate any underlying host galaxy light which is present in both spectra and reveal the SED of the variable component. 
Aside from the features resulting from the BEL change, the  $|\Delta f_{\lambda}|$ curves are well-described by a power-law of the form $f_{\nu}\propto \nu^{\beta}$, and the best-fit power-law is shown as a blue curve.  The median best-fit power-law index for the sample is $\beta=0.2$, with an root-mean-square value 0.4.
The red dashed curves show a $f_{\nu}\propto \nu^{1/3}$ power law expected for a standard thin disk, and the difference spectra for all 10 objects appear qualitatively consistent with this form.  
Note that in some cases, $|\Delta f_{\lambda}|$ deviates from the power-law at the shortest wavelengths (e.g., top two panels of Fig.~\ref{fig:disappear}), but this is likely due to a relatively increased noise level in the SDSS spectrum. 

It is well-known that AGN spectra do not typically follow the $f_{\nu}\propto \nu^{1/3}$ form expected for a standard thin disk in the optical; only the NIR polarized continuum has been shown to agree with this prediction \citep{kis03}. This is true for our final sample as well (the median $\beta\approx-1$ in the high-luminosity state).  However,  $|\Delta f_{\lambda}|$ for all 10 objects follows a power-law that is much more consistent with  $f_{\nu}\propto \nu^{1/3}$ than the individual spectra.  This implies that the variable component is related to the thermal accretion disk, which might be expected if there was a change in the temperature structure of the accretion disk due to a change in viscous heating or irradiating flux.

 \subsection{Comparison to Previously Reported Changing-Look Quasars}

In Table~\ref{tab:prev_CLQ}, we list four previously reported changing-look quasars (at $z>0.1$), including J0159+0033. J012648.08$-$083948.0 and J233602.98$+$001728.7 are not in DR7Q and therefore are not in our parent sample.  These two objects were found by \citet{rua15} in an archival search for objects whose pipeline spectroscopic classifications changed between epochs of spectroscopy. Therefore, their search was sensitive to only the most dramatic changes in spectral state, whereas our variability-based selection is sensitive to dramatic transitions as well as more subtle changes in AGN type. However, while J233602.98$+$001728.7 shows enough variability to be included in our sample, J012648.08$-$083948.0 only shows a 0.78~mag dimming going from SDSS to PS1.  This suggests that our variability criterion of $|\Delta g|>1.0$~mag is too restrictive. 

The \citet{rua15} sample selects objects that change their "{\tt CLASS}" classification,  
where the {\tt CLASS} parameter is from the 1-D SDSS pipeline \citep[see Section 5,][]{bol12}.  
All the objects in our sample except for two are originally classified 
as ``{\tt QSO}'' {\it and stay} classified as {\tt CLASS}$=${\tt QSO} in their later epoch pipeline classification. 
The two exceptions are J0159+0033 and J132457.29+480241.2. However, the 
latter source was not considered in the end to be a changing-look quasar by Ruan et al.\ due 
to reasons described in the Appendix of that paper. 

The Time Domain Spectroscopic Survey of SDSS-IV \citep[TDSS;][]{mor15} is obtaining repeat spectra for quasars showing $|\Delta g|>0.7$~mag variability, and is likely to find a more complete sample of changing-look quasars. 
J101152.98$+$544206.4, discovered by TDSS \citep{run15}, is in DR7Q but does not have a spectrum in DR12. Had it been observed in SDSS-III, we would have recovered this object since it is $1.5$~mag dimmer in PS1 ($g=19.879\pm 0.014$ at ${\rm MJD}=55838.170$) with respect to SDSS.  

\begin{table}
  \begin{center}
    \begin{tabular}{lcccc}
      \hline
      \hline
Name     & $z$ & Max($|\Delta g|$) & DR7$\phantom{0}$  & Found  \\
(SDSS J) &     & (mag)           &   QSO? & here?   \\
      \hline
012648.08$-$083948.0$^a$  & 0.198 & $0.78 \pm 0.03$ & $\times$     & $\times$     \\
015957.64$+$003310.4$^b$  & 0.312 & $1.16 \pm 0.06$ & $\checkmark$ & $\checkmark$ \\
101152.98$+$544206.4$^c$  & 0.246 & $1.51 \pm 0.02$ & $\checkmark$ & $\times$     \\
233602.98$+$001728.7$^a$  & 0.243 & $1.59 \pm 0.13$ & $\times$     & $\times$     \\
      \hline
      \hline
    \end{tabular}
    \caption{Previously reported changing-look quasars ($z>0.1$).
$^a$Found by \citet{rua15}.
$^b$Found by \citet{lam15}.
$^c$Found by \citet{run15}.
}
    \label{tab:prev_CLQ}
  \end{center}
\end{table}


\section{Discussion}\label{disc}

AGN are known to be variable phenomena. However, it is only relatively recently that
multiple spectra of the same AGN at high-$z$ have become available, mainly
due to large spectroscopic surveys such as the SDSS. Moreover, noting {\it changes} 
in these spectra of AGN gives direct observational evidence and insight to the 
physical processes that are happening in the AGN.

One key result from our  systematic search is that using a variability cut seems to be one effective way of finding changing-look quasars.  Our base sample is DR7Q, of which 6348 (6\%) satisfy the photometric selection described in Section~\ref{sample}. Of these, 1011 objects are reobserved with BOSS, and 10 (1\%) show BEL (dis)appearances and reside in the redshift range  $0.20<z<0.63$.  We may be biased toward finding changing-look behavior at $z<0.8$ since beyond this the H$\beta$ line is no longer in the SDSS spectrum, and significant changes were most commonly seen in this line. The lack of significant changes in the \civ and \mgii lines may be due to the fact that \civ is affected by winds \citep{ric11} and that \mgii has a relatively weak responsivity \citep{goa93,cac15}. 

A second key result from our study is that the difference spectra for our final sample of 10 changing-look quasars are more consistent with the naively expected $f_{\nu}\propto \nu^{1/3}$ power-law for a thin disk \citep{SS73} than the individual spectra. This suggests that the variable component has an SED similar to an accretion disk, and that the reddening in the quasar host galaxy must be small. Furthermore, whatever is causing a BEL (dis)appearance must be linked to an emerging (diminishing) continuum. Indeed, the BELs track the blue continuum change on rest-frame timescales as short as 255~days in J0225+0030 (Section~\ref{both}).

``Outbursts'' associated with BEL appearance could be due to changing obscuration, changing accretion rate, or transient behavior such as a tidal disruption event (TDE) or microlensing. Presumably because of the relatively low redshifts, these are not microlensing events \citep[see][for larger amplitude and typically more distant AGN flares]{law12b}.  While a TDE could explain J0159+0033 \citep{mer15}, such a flaring episode is ruled out for at least some other objects presented here.  For example, the light curve in the top panel of Fig.~\ref{fig:appear} shows that the luminosity has remained in the high state for the last 2000 days (observed frame) instead of decaying with a $t^{-5/3}$ form expected for a TDE \citep[e.g.,][]{eva89,gez12,gui14}. 
Furthermore, TDEs should only remain in the high state for 
a few months \citep[e.g.,][]{gez12}, unlike what is observed here.  In addition, the narrow 
emission lines are present in the early states for each source, which would be too fast of a large-scale response 
for a TDE \citep[as pointed out by][]{run15,rua15}.
Therefore, we limit the following discussion to the first two possible scenarios.

\subsection{Observed Timescales for BEL (Dis)appearance}
\label{timescales}
\begin{figure}
\centering
\includegraphics[scale=.48]{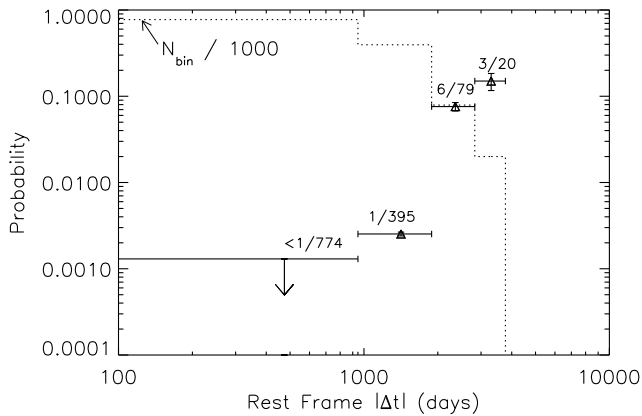}
\caption{Probability, $p$, of BEL (dis)appearance as a function of rest-frame time lag, based on our sample (see text). The errors in $p$ are computed as $p/\sqrt{N_{\rm bin}}$, where   $N_{\rm bin}$  (shown as the normalized, dotted histogram) is the number of highly variable quasars with repeat spectra (red open data points in Fig.~\ref{fig:dt_dg}) which fall in each time bin.  None of our 10 objects have $|\Delta t|<1000$~days, so for this bin we only show an upper limit corresponding to $<1$~object. The ratio of changing-look objects to $N_{\rm bin}$ is listed in each bin. } 
\label{fig:prob}
\end{figure}

One feature which  may help guide physical interpretations is the timescale associated with a spectral state change.
As such, we would like to place an upper limit on the \emph{fastest} timescales associated with BEL (dis)appearance. Due to the nature of the SDSS and PS1, we lack a complete sampling of timescales and are prone to selection effects imposed by the timing of both the photometric and (especially the) spectroscopic observations. In Fig.~\ref{fig:prob}, we show the probablility $p$ which is simply the ratio of changing-look quasars to the highly variable sample with repeat spectra, i.e., the ratio of blue crosses to red open circles in each of the four time bins shown in the bottom panel of Fig.~\ref{fig:dt_dg}. Out of the highly variable objects with repeat spectra, we find the highest probability for changing-look behavior on rest-frame timescales of 2000--3000 days and 3000--4000 days ($p=0.076\pm 0.009$ and $p=0.15\pm 0.03$, respectively). When restricted to the redshift range $0.2<z<0.63$, $p$ increases to $p=0.113\pm 0.016$ and $p=0.18\pm 0.04$, respectively.  
Note that $p$ will be higher for the subsample with \emph{spectroscopic} epochs spanning a $>1$~mag change (i.e., those showing a $>1$~mag change from one spectral epoch to the next, shown by the red open circles above the dashed line in Fig.~\ref{fig:dt_dg}) -- this subsample is more representative of the objects in which we could have observed changing-look behavior. However, the size of this subsample is too small to make any statistical conclusions. 
Along with being biased by the sparse sampling of spectroscopic epochs, we are also likely biased toward long rest-frame timescales given that H$\beta$ is only visible at $z<0.8$. Fig.~\ref{fig:prob} suggests that the fraction of changing-look quasars rises to 0.2 at $\Delta t\sim 10$~yr; therefore, future surveys are needed to determine if the fraction of changing-look quasars continues to rise on even longer timescales.

While the timescales for BEL changes explored in Fig.~\ref{fig:prob} are limited to the timing of spectroscopic observations, seven of our objects have S82 light curves, which provide more information on the timescales over which the transitions may have occurred. In general, the light curves show strong increases in flux over 1000 days in the observed frame (or roughly two years in the quasar rest frame; Fig.~\ref{fig:appear}), but strong decreases in flux over considerably longer timescales (several years in the observed frame; top two panels of Fig.~\ref{fig:disappear}).  The light curve for J0225+0030 (Fig.~\ref{fig:both}) shows a rise over less than a year in the quasar rest frame following a slow decline over 5 years in the rest frame.

\subsection{A Change in the Central Engine?}

To explore a physical scenario where the amount of available ionizing flux from the central engine has changed, we must consider both the timescale for BEL response as well as the timescale for the continuum variability. 
In the first case, since BELs result from photoionization \citep[e.g.,][]{pet93}, they should respond on the light crossing timescale $t_{lt} = R_{\rm BLR}/c$~days, where $R_{\rm BLR}$ is the radius of the BLR.  Using the $R-L$ relation calibrated by \citet{ben13}, the BLR size is estimated to be $R_{\rm BLR}=2954 R_S M_8^{-1} L_{44}^{0.533}$, where $M_8=M_{\rm BH}/(10^8 {\rm M}_{\odot})$ is the mass of the central super-massive black hole in units of $10^8 {\rm M}_{\odot}$, $L_{44}=\lambda L_{\lambda}(5100{\rm \AA}) / (10^{44}~{\rm erg~s^{-1}})$, and $R_S$ is the Schwarszchild radius $R_S=2GM_{\rm BH}/c^2$.  This gives $t_{lt}=34 L_{44}^{0.533}$~days, similar to the observed BEL lags in reverberation mapping (RM), although the lags are typically shorter since the majority of the $\sim$50 AGN studied through RM have lower luminosities.

One assumption in RM studies is that the structure of the BLR remains stable over the duration of the experiment. For Seyferts, the optical continuum variations are typically a factor 1.3 over rest-frame timescales of $\sim$months \citep[e.g.,][]{ede15}.  The photometric variability presented here is more dramatic in comparison:  on average by a factor 4 in $g$-band flux over seven years in the rest frame.
Furthermore, we observe a stronger  \mgii  response in J0225+0030 over 4.5 years in the rest frame (Section~\ref{both}) than typical in RM studies \citep[e.g.,][]{cac15}, which might be expected if the source of ionizing photons has significantly diminished.   
In this case, the BLR may have time to adjust its overall structure in response to such large changes in ionizing flux, and BEL changes might be expected on the dynamical timescale of the BLR.  For typical Seyfert galaxies,  $t_{\rm dyn}\approx R_{\rm BLR}/\Delta V \approx 3$~to 5~years \citep{pet06}, where $\Delta V$ is a typical cloud velocity.  This timescale will be slightly longer for higher-luminosity quasars, since $t_{\rm dyn}\propto L^{3/4} M_{\rm BH}^{-1/2}$, assuming Keplerian rotation and that $R_{\rm BLR} \propto L^{1/2}$ for photoionized lines. The time between SDSS and BOSS spectra is long enough so that a dynamical response of the BLR cannot generally be ruled out for our sample.

Regardless of what is happening in the broad line region, it is clear that the BELs track a large change in continuum level flux.
The timescale that might be associated with an accretion rate change is the viscous, or ``radial inflow'' timescale \citep[see e.g.,][]{kro99}. 
Indeed, \citet{eli14} provide a scenario where AGN evolve naturally from Type 1 to 1.2/1.5 to 1.8/1.9 as the accretion rate diminishes.  
Using Equation~5 in \citet{lam15} and scaling the Eddington parameter $\lambda_{\rm Edd}$ and $M_{\rm BH}$ to the measured values for J1021+4645 from \citet{she11}, we obtain:
\begin{equation}
  t_{\rm infl} = 5\times 10^4 \left[ \frac{\alpha}{0.1} \right]^{-1}  
                         \left[ \frac{\lambda_{\rm Edd}}{0.05} \right]^{-2}  
                        \left[ \frac{\eta}{0.1} \right]^{2}  
                        \left[ \frac{r}{50 R_{\rm S}} \right]^{7/2}  
                        \left[ \frac{M_{8}}{2.1} \right] {\rm yr.}
\end{equation}
Here, $\alpha$ is the disk viscosity parameter, $\eta$ is the accretion efficiency, and $r$ is 
the accretion disk radius (assumed to be $50R_{\rm S}$ for optical disk emission). 
The value of $t_{\rm infl}$ may be a several times shorter based on magneto-hydrodynamical simulations \citep[e.g.,][]{kro05}, but this is still too long to explain the continuum variability of all sources presented here (see the Appendix for the individual physical parameters and $t_{\rm infl}$ estimates). The source J022652.24$-$003916.5 has a relatively low mass and therefore is expected to have a relatively short $t_{\rm infl}$ of $\sim$56~yr, still an order of magnitude too long to explain the observed optical variability. Furthermore, \citet{lam15} point out that  $t_{\rm infl}$ should be considerably longer in the dim state since $t_{\rm infl}\propto \lambda_{\rm Edd}^{-2}$, and thus the dimming timescale should be shorter than the recovery timescale.   However, the S82 light curves suggest the opposite behavior: the objects with appearing BELs show brightening on much shorter timescales than is plausible for the optical viscous timescale in the dim states of these AGN. 

Nevertheless,  $t_{\rm infl}$ might still be an appropriate timescale if the optical flux contains reprocessed emission from the EUV, where  $t_{\rm infl}$ may be orders of magnitude shorter.  In the disk reprocessing scenario, the disk surface is irradiated by EUV/X-ray photons coming from a central source on the light travel timescale, and the re-emitted UV/optical flux can vary on the shorter timescales associated with the central region. This ``lamp post'' model \citep[e.g.,][]{ber00,mar02} is supported by observations of quasar microlensing \citep[e.g.,][]{mor08,mos13,mac15}, which show that the X-rays originate from projected radii near the inner edge of the accretion disk. Indeed, multi-wavelength observations of NGC~4051 indicate that $\sim$25\% of the UV variance is caused by thermal reprocessing \citep{als13}. In the disk reprocessing model of \citet{cac07}, the disk temperature $T_0$ at a fiducial radius $R_0$ is regulated by both local viscous heating and irradiation:
\begin{equation}
      T_0^4 = T_{\rm infl}^4 + T_{\rm irr}^4 = \frac{3 G M_{\rm BH} \dot{M} }{ 8 \pi \sigma R_0^3} + \frac{L_X (1-A) h }{4 \pi \sigma R_0^3},
\end{equation}
where $L_X$ is the lamp post luminosity at a height $h$ above the black hole, and $A$ is the disk albedo. In this case, the temperature-radius profile of the disk would rise and fall with changes in either local accretion rate or irradiation, and this would produce a difference spectrum that remains an $f_{\nu} \propto \nu^{1/3}$ power-law \citep{cac07}, similar to what we observe in the objects presented here. Note that the fraction of reprocessed flux that is re-emitted toward the interior of the disk can modify its internal structure, but this effect is negligible unless $L_X$ is similar to the bolometric luminosity of the quasar \citep{col01}. 

\subsection{Variable Obscuration?}

In the case of variable extinction, a passing cloud may obscure the BLR on a characteristic crossing timescale $t_{\rm cross}$. In the simplest scenario, such an event must account for the disappearing BELs and thermal continuum. One possibility is a passing isolated BLR cloud obscuring the continuum emitting region as seen by us and by the (remainder of the) BLR. While this scenario is viable in the X-rays where the continuum emitting source is relatively compact \citep[e.g.,][]{ris09}, the UV/optical continuum emitting region being significantly larger makes this scenario less likely. But the timescale for such a crossing event is reasonable:  $t_{\rm cross, BLR~cloud}=84 M_8^{-1/2} L_{44}^{3/4}$~days assuming a Keplerian orbit for a BLR cloud passing at $r_{\rm orb}=R_{\rm BLR}=2954 R_S M_8^{-1} L_{44}^{1/2}$ in front of the continuum source, where the latter has a radius $r_{\rm src}= R_{\rm BLR}/60$ \citep[see Equation~4 in][]{lam15}. 

Another possibility is that the entire UV/continuum emitting region and BLR is obscured; this might be expected if the distribution of dust lying at the outskirts of the BLR is patchy \citep[e.g.,][]{nen08a,nen08b,eli12}. In this case, reasonable values would be $r_{\rm src}=R_{\rm BLR}=2954 R_S M_8^{-1} L_{44}^{1/2}$ and $r_{\rm orb}=3R_{\rm BLR}$ for dust at $r_{\rm orb}$ to obscure a significant portion of the BLR \citep{lam15}. This gives $t_{\rm cross,dust}= 24 M_8^{-1/2} L_{44}^{3/4}$~yr.

The $t_{\rm cross}$ estimates are listed in Table~\ref{tab:app} for each source.  The $t_{\rm cross, BLR~cloud}$ estimates are shorter than the observed brightening and dimming timescales for all sources, but since the size and structure of BLR clouds is uncertain, these timescales may be longer than predicted. The $t_{\rm cross,dust}$ values are too long to explain the light curve behavior of these sources but are in many cases of the same order as the observed timescales. For example, the source SDSS J233317.38-002303.4 has the shortest $t_{\rm cross,dust}$ estimate of 4.5 yrs (based on the luminosity in the bright state), but the observed brightening happens over 2.7~yr in the rest frame (top panel of Fig.~\ref{fig:appear}). For the densely monitored object J0225+0030, the predicted dust crossing timescale is an order of magnitude too long to explain the rebrightening around MJD$=55500$ (Fig.~\ref{fig:both}).  Nevertheless, we further investigate the plausibility of variable obscuration by considering simple dust reddening models.

\subsubsection{Dust Reddening Test}
\label{1021}
 We consider whether an obscuration event can account for the spectral changes by applying dust reddening to the observed spectra. In particular, we simply apply Milky Way (MW) and Small Magellanic Cloud (SMC) dust reddening curves to the bright-state spectra and see whether we can reproduce the BEL disappearance in the faint-state spectra. This test is demonstrated here for SDSS J1021+4645, which has high quality spectra that show the most dramatic transition from a Type 1.0 AGN in SDSS to a Type 1.9 AGN in BOSS among our objects. 

\begin{figure*}
\centering
\includegraphics[scale=.65]{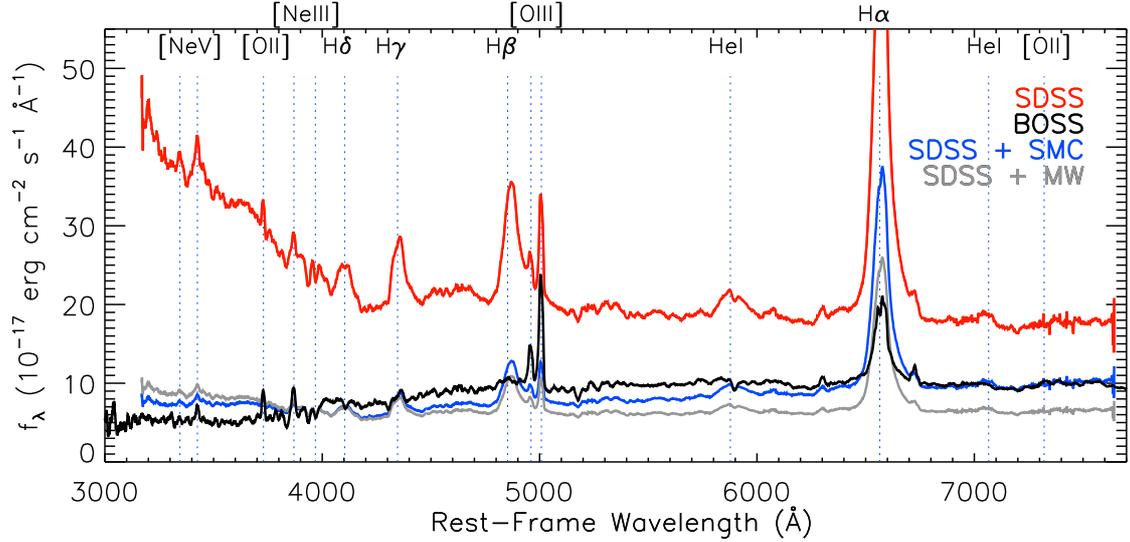}  
\caption{SDSS~J1021+4645, showing a dramatic transition from a Type 1.0 AGN in SDSS (shown in red) to a Type 1.9 AGN in BOSS (shown in thick black). The red spectrum shows the observed SDSS flux; the grey and blue spectra show  extinguished versions of the SDSS spectrum using a MW and an SMC extinction curve, respectively (see text for details). Applying reddening to the SDSS spectrum is inadequate for reproducing the BOSS spectrum, such that a simple extinction scenario can be ruled out as an explanation for the transition.}
\label{fig:1021}
\end{figure*}

Assuming the SDSS spectrum for SDSS J1021+4645 is the intrinsic, unabsorbed flux, we estimate $A_V=0.95$  by taking the magnitude difference between the SDSS and BOSS spectra at the effective wavelength of the Johnson $V$ band.  The SDSS spectrum is then multiplied by a factor $10^{-A_{\lambda}/2.5}$, where $A_{\lambda}=E(\lambda-V) + A_V$, $E(\lambda-V)=k(\lambda) E(B-V)$, and $k(\lambda)$ is an extinction curve. For SMC-like extinction, we tried the four different models in Table~5 of \citet{gor98}, which use the parametrization in  Equation~2 of \citet{fit90}. We find that the AzV 214 model ($R_V=2.75$) produces an absorbed spectrum most similar to the observed BOSS spectrum among the four fits, and the result is shown in Fig.~\ref{fig:1021}.  We also try a MW-like extinction curve \citep[$R_V=3.1$;][]{sea79}, and assume $A_V=0.95$ as before. Note that the SDSS spectrum will contain a host galaxy component that is not modeled here. However, we are not interested in reproducing the detailed continuum shape of the BOSS spectrum but rather the BEL change, so we leave a detailed spectral modeling for a future study as it is should not affect our final conclusions.

There are two points to note from the simple extinction tests in Fig.~\ref{fig:1021}. First, the SMC-like extinction curve does a better job than the MW-like extinction curve at removing the blue part of the continuum in the SDSS spectrum.  Second, while the SMC-like extinction model can qualitatively explain the continuum change going from SDSS to BOSS, neither dust model can explain the disappearance of the BELs. In particular, the H$\beta$ BEL remains prominent in the absorbed versions of the SDSS spectrum.  This latter point is similar to what \citet{lam15} find for J0159+0033, where the observed H$\alpha$ BEL is much stronger than it should be if reddening alone is responsible for the spectral change. \citet{lam15} also find that archival X-ray spectra in the bright and faint states of  J0159+0033 both appear unabsorbed. 

We reach the same conclusion when repeating the above exercise for the other nine sources, using each pair of spectra listed in Table~\ref{tab:objs}: if reddening alone could account for the observed spectral changes, then the broad component of H$\beta$  should be present in the faint-state spectrum on top of the host galaxy flux. However, the broad line flux is instead negligible in each case.    
Therefore, simple dust reddening models fail to simultaneously account for the diminished continuum and broad Balmer line components in all ten sources. In principle, one could employ a different $A_V$ for the BELs than for the continuum to explain the optical spectral change, but this leads to a more complicated physical scenario than we explore here.

\section{Conclusions}
\label{conc}
We have presented the first systematic search for changing-look quasars using data from the SDSS and PS1 3$\pi$. 
Out of 6348 quasars from DR7Q that exhibit $|\Delta g|>1$~mag variability over the SDSS and PS1 photometric epochs, 1011 have been reobserved with BOSS.  Of these, we visually identify 10 changing-look quasars, extending the known population of changing-look AGN to $z=0.63$.

Our search has yielded the following:
\begin{itemize}
    \item{four objects with appearing BELs corresponding to $>1$~mag increases in flux in the $g$-band;}
    \item{five objects with disappearing BELs and declining light curves, which includes the changing-look 
      quasar discovered in \citet{lam15}, and an even more dramatic transition from a Type 1.0 to 1.9 AGN 
      in the object J1021+4645 (with a flux deviation per spectral pixel of $8\sigma$ for H$\beta$);}
    \item{one object showing evidence of both disappearing and appearing BELs on timescales less than a year in the quasar frame;}
    \item{an SED of the variable component resembling thermal emission from an accretion disk;}
    \item{that $\sim 15$\% of quasars that have varied by  $|\Delta g|>1$~mag at some point in their 
      light curve display changing-look behavior on rest-frame timescales of 3000--4000~days;}
    \item{that a change in ionizing flux from the central engine coupled with disk reprocessing is a more 
      likely explanation than variable obscuration in all cases, based on timescale arguments and simple 
       dust reddening models.}
\end{itemize}

While the detailed physical mechanisms behind BEL changes may be complicated to
understand and model, changing-look AGN can provide a laboratory to study the
relationship between emission at different wavelengths, as the entire
system may respond to a strong increase or decline in available ionizing flux. 
As a prime example, following an outburst in the X-ray emission in the local
Seyfert NGC 2617, \citet{sha14} were able to use the variability
across several wavebands (spanning X-rays--NIR) to map out the
structure of the accretion disk and argue for illumination of the disk by the
X-ray source. The observations presented here support this scenario, 
although we cannot rule out a change in obscuration without more complex models 
for the structure and size of the intervening material.

Our results indicate that photometric variability might be one of the 
best ways to efficiently find changing-look quasars in current datasets. Furthermore, 
with well-sampled light curves provided by the S82 survey, extreme behavior 
can be more readily identified, as demonstrated by the higher yield of 
changing-look objects from S82 presented here.
Therefore, future time domain surveys such as the Large Synoptic Survey Telescope 
\citep{ive08,lsst}, PS1 and PS2 \citep{cha14}, as well as the ongoing  All-Sky Automated Survey for SuperNovae
\citep{sha14}, should provide excellent datasets in which to search for 
interesting targets for spectroscopic follow-up. 
Based on the observations in \citet{sha14} and \citet{lam15}, extreme X-ray
variability may also help identify changing-look AGN candidates 
among samples of extremely optically-variable quasars. Furthermore, a 
multi-wavelength monitoring campaign of these objects would be useful
for interpreting the nature of such transitions.  However, since it is
unclear how rare these transitions are at any given timescale, further work is necessary for
determining the best strategy for follow-up monitoring.  The TDSS project of SDSS-IV \citep{mor15} has already initiated 
a similar search to the one presented here, extending the search criteria down to 
$\Delta g>0.7$~mag, and should place more stringent limits on
the frequency of such events. On the longer term, the Dark Energy
Spectroscopic Instrument (DESI) will survey $\sim 2.4\times 10^6$
quasars \citep{lev13}, providing ample opportunity for obtaining additional
spectroscopic epochs for highly-variable AGN.

\section*{Acknowledgements}

We acknowledge Marco Lam and Nigel Hambly for assistance with and maintaining the local PS1 DVO database. 
We also acknowledge Isabelle P\^{a}ris for help with checking which of our objects were in the SDSS-III BOSS Quasar catalog. We thank the reviewer for valuable suggestions that improved the paper. 
CLM acknowledges support from the STFC Consolidated Grant (Ref.\ St/M001229/1). NPR acknowledges support from the STFC and the Ernest Rutherford Fellowship scheme. KH acknowledges support from STFC grant ST/M001296/1.

    Funding for the SDSS and SDSS-II has been provided by the Alfred P. Sloan Foundation, the Participating Institutions, the National Science Foundation, the U.S. Department of Energy, the National Aeronautics and Space Administration, the Japanese Monbukagakusho, the Max Planck Society, and the Higher Education Funding Council for England. The SDSS Web Site is http://www.sdss.org/.

    The SDSS is managed by the Astrophysical Research Consortium for the Participating Institutions. The Participating Institutions are the American Museum of Natural History, Astrophysical Institute Potsdam, University of Basel, University of Cambridge, Case Western Reserve University, University of Chicago, Drexel University, Fermilab, the Institute for Advanced Study, the Japan Participation Group, Johns Hopkins University, the Joint Institute for Nuclear Astrophysics, the Kavli Institute for Particle Astrophysics and Cosmology, the Korean Scientist Group, the Chinese Academy of Sciences (LAMOST), Los Alamos National Laboratory, the Max-Planck-Institute for Astronomy (MPIA), the Max-Planck-Institute for Astrophysics (MPA), New Mexico State University, Ohio State University, University of Pittsburgh, University of Portsmouth, Princeton University, the United States Naval Observatory, and the University of Washington.

Funding for SDSS-III has been provided by the Alfred P. Sloan Foundation, the Participating Institutions, the National Science Foundation, and the U.S. Department of Energy Office of Science. The SDSS-III web site is http://www.sdss3.org/.

SDSS-III is managed by the Astrophysical Research Consortium for the Participating Institutions of the SDSS-III Collaboration including the University of Arizona, the Brazilian Participation Group, Brookhaven National Laboratory, Carnegie Mellon University, University of Florida, the French Participation Group, the German Participation Group, Harvard University, the Instituto de Astrofisica de Canarias, the Michigan State/Notre Dame/JINA Participation Group, Johns Hopkins University, Lawrence Berkeley National Laboratory, Max Planck Institute for Astrophysics, Max Planck Institute for Extraterrestrial Physics, New Mexico State University, New York University, Ohio State University, Pennsylvania State University, University of Portsmouth, Princeton University, the Spanish Participation Group, University of Tokyo, University of Utah, Vanderbilt University, University of Virginia, University of Washington, and Yale University.

The Pan-STARRS1 Surveys (PS1) have been made possible through contributions of the Institute for Astronomy, the University of Hawaii, the Pan-STARRS Project Office, the Max-Planck Society and its participating institutes, the Max Planck Institute for Astronomy, Heidelberg and the Max Planck Institute for Extraterrestrial Physics, Garching, The Johns Hopkins University, Durham University, the University of Edinburgh, Queen's University Belfast, the Harvard-Smithsonian Center for Astrophysics, the Las Cumbres Observatory Global Telescope Network Incorporated, the National Central University of Taiwan, the Space Telescope Science Institute, the National Aeronautics and Space Administration under Grant No. NNX08AR22G issued through the Planetary Science Division of the NASA Science Mission Directorate, the National Science Foundation under Grant No. AST-1238877, the University of Maryland, and Eotvos Lorand University (ELTE).

The CSS survey is funded by the National Aeronautics and Space
Administration under Grant No. NNG05GF22G issued through the Science
Mission Directorate Near-Earth Objects Observations Program.  The CRTS
survey is supported by the U.S.~National Science Foundation under
grants AST-0909182 and AST-1313422.

This research has made use of data obtained from the Chandra Source Catalog, provided by the Chandra X-ray Center (CXC) as part of the Chandra Data Archive.

{\it Facilities:} SDSS; BOSS; Pan-STARRS1

%
%




\bibliographystyle{mnras}
\bibliography{refs} 




\appendix
\section{Notes on Individual Objects}
A list of physical properties for the final sample is provided in Table~\ref{tab:app}.
The luminosities, black hole masses, and Eddington ratios listed are the original SDSS DR7 measurements from \citet{she11}.  We adopt these values when calculating $R_{\rm BLR}$ and the viscous timescale $t_{\rm infl}$, since the original luminosity will determine the timescale for subsequent dimming or recovery. However, for the crossing timescales, we adopt the luminosity at 5100\AA\ in the bright state, since this luminosity is presumed to be the intrinsic value in an obscuration scenario. Therefore, for the four quasars in Figure~\ref{fig:appear} that have ``turned on'', the DR7 $L_{5100}$ values are multiplied by factors of 1.3--1.8 as appropriate when determining $t_{\rm cross}$.  

\begin{table*}
  \caption{Properties of changing-look quasars,  
    along with estimates for viscous timescale, $t_{\rm infl}$, and crossing timescales for a BLR cloud and for dust ($t_{\rm cross, BLR~cloud}$ and $t_{\rm cross, dust}$, respectively).
    Masses, luminosities, and Eddington ratios are taken from \citet{she11}.}
  \label{tab:app}
  \begin{tabular}{l cccc ccc}
    \hline
    \hline
    Name & $\log M_{\rm BH}/M_{\odot}$ & $\log L_{5100} $ & $\log L/L_{\rm Edd} $ & $R_{\rm BLR}/R_{\rm S}$ & $t_{\rm infl}$ &  $t_{\rm cross, BLR~cloud}$&  $t_{\rm cross, dust}$\\
  (SDSSJ)    &    & $[{\rm erg~s^{-1}}]$ &   &   & (yr)  & (days)  & (yr)\\
    \hline
  002311.06$+$003517.5 &  9.23 & 44.51 & -1.85 &   329 &  4.4E+06 &  67.6 &  19.3\\
  015957.64$+$003310.4 &  8.40 & 44.21 & -1.32 &  1532 &  5.8E+04 &  76.5 &  21.9\\
  022556.07$+$003026.7 &  8.35 & 44.38 & -1.10 &  2112 &  1.9E+04 & 108.9 &  31.1\\
  022652.24$-$003916.5 &  7.48 & 44.34 & -0.27 & 14971 &  5.6E+01 & 275.4 &  78.7\\
  100220.17$+$450927.3 &  8.96 & 44.45 & -1.64 &   569 &  9.1E+05 &  61.1 &  17.5\\
  102152.34$+$464515.6 &  8.33 & 44.15 & -1.31 &  1664 &  4.8E+04 &  74.9 &  21.4\\
  132457.29$+$480241.2 &  8.51 & 44.34 & -1.30 &  1394 &  6.9E+04 &  83.9 &  24.0\\
  214613.31$+$000930.8 &  8.94 & 44.25 & -1.82 &   464 &  2.0E+06 &  52.9 &  15.1\\
  225240.37$+$010958.7 &  8.88 & 44.35 & -1.66 &   602 &  8.4E+05 &  85.7 &  24.5\\
  233317.38$-$002303.4 & 10.15 & 44.31 & -2.97 &    30 &  6.6E+09 &  15.6 &   4.5\\
    \hline
    \hline
  \end{tabular}
\end{table*}

    \subsection{SDSS J002311.06+003517.5}
    {\tt SN\_GAL1} target in BOSS spectroscopy, indicating this 
    object was associated with obtaining the spectrum of host 
    galaxies of SDSS-II supernovae \citep{Dawson13}. 
    This object is in the Supplementary List of the 
    SDSS-III BOSS DR12 Quasar catalog (DR12Q;  {P{\^a}ris} et al.~2016, in advanced prep.) with $z = 0.422$.
    
    \subsection{SDSS J015957.64+003310.4}
    {\tt TEMPLATE\_QSO\_SDSS1} target flag in BOSS. 
    This object is not in DR12Q since it is flagged 
    as a ``Starforming galaxy'' by the BOSS 1-D spectral classification 
    pipeline. 

    \subsection{SDSS J022556.07+003026.7}
    \label{0225}
    {\tt TEMPLATE\_GAL\_PHOTO} target flag in BOSS. 
    This object is in the DR12Q Supplementary list with $z = 0.504$.
    
    Its S82 light curve and multiple BOSS spectra show a 1.5 magnitude dimming over 4.5~yr (rest-frame), during which H$\beta$ vanishes and \mgii strongly diminishes, and then a rebrightening during which the broad components of the Balmer lines (re)appear within 0.7~yr in the rest frame.  When measuring the \mgii EW, we estimate the local continuum flux $f_{0,\lambda}(\lambda)$ by subtracting a Gaussian fit to the \mgii line from the observed flux $f_{\lambda}(\lambda)$ and integrating the quantity $(1 - [f_{\lambda}(\lambda)/f_{0,\lambda}(\lambda)])$ over the line. 

    We find a constant \mgii EW over the 1.5 magnitude dimming from MJD$=$52944 to 55445, indicating a linear 
    response to the change in flux and a negligible Baldwin effect. What is typically observed for an individual 
    object (with lower-amplitude variability) is an unresponsive MgII line \citep[e.g.,][]{cac15}, so our 
    results suggest that the source of ionizing photons had significantly diminished during this period.

    We find from our test in (Section~\ref{1021}) that extinction by dust can easily explain the large change in the continuum flux and \mgii line, but not the change in H$\beta$. Compared to the observed faint-state spectrum, there is an excess of H$\beta$ flux in the predicted extinguished spectrum at marginal significance ($\sim 1\sigma$ per spectral pixel). Furthermore, the predicted dust crossing timescale for J0225+0030 is about four times too long to explain the dimming in the light curve and $\sim$10 times too long to explain the subsequent rebrightening.

    \subsection{SDSS J022652.24-003916.5}
    {\tt QSO\_VAR\_LF} target flag in BOSS indicating this 
    object was specifically targeted due to its variable nature 
    of quasar luminosity function studies 
    \citep{Palanque_Delabrouille13}. 
    This object is in DR12Q. 
    The relatively small black hole mass for this object 
    yields a viscous timescale $t_{infl}$ that is most similar to the observed 
    optical dimming timescale of several years, but is still an order of magnitude too long (see Table~\ref{tab:app}). 

    \subsection{SDSS J100220.17+450927.3}
    J100220.17+450927.3 was selected as being a {\tt  QSO\_CAP} and a {\tt ROSAT\_B}, 
    {\tt ROSAT\_D} and {\tt ROSAT\_E} target in SDSS \citep{ric02},  
    and then was selected to be observed in SDSS-III BOSS via the {\tt SEQUELS\_TARGET} and 
    {\tt TDSS\_FES\_DE} target flags \citep[{P{\^a}ris} et al.~2016, in advanced prep.;][]{Myers15}. 

    SEQUELS is the Sloan Extended QUasar, ELG and LRG Survey, undertaken as part of SDSS-III 
    in order to prepare and refine target selection for the SDSS-IV extended Baryon Oscillation Spectroscopic Survey
    \citep[eBOSS;][]{Dawson15}. Another part of SDSS-IV is the Time Domain Spectroscopic Survey \citep[TDSS;][]{mor15} 
    and the {\tt TDSS\_FES\_DE} flag is set for quasar disk emitters selected to be observed by TDSS. These targets are quasars 
    with $i < 18.9$ and broad, double-peaked or asymmetric Balmer emission line profiles, such as those in 
    \citet[][$z<0.33$ for H$\alpha$ and H$\beta$]{str06} and higher-redshift analogs from 
    \citet[][$z\sim 0.6$ for H$\beta$ and \mgii]{luo13}. 
    This object appears in DR12Q. For a description of the SDSS-III SEQUELS target flags, see \citet{ala15}.

    \subsection{J102152.34+464515.6}
    J1021 was selected via the {\tt QSO\_CAP} target flag in SDSS, and then via 
    {\tt SEQUELS\_TARGET} and {\tt TDSS\_FES\_NQHISN} in BOSS.
    The {\tt TDSS\_FES\_NQHISN} target flag is set for $z<0.8$ DR7 quasars with 
    high signal-to-noise spectra to study broad-line variability 
    on multi-year timescales \citep{ala15}. 

    \subsection{SDSS J132457.29+480241.2}
    {\tt SEQUELS\_TARGET} and {\tt TDSS\_FES\_NQHISN} target in BOSS.
    It is not in the DR12Q because of the missing red spectrum. 

    \subsection{SDSS J214613.31+000930.8}
    {\tt SN\_GAL1} target flag in BOSS. 
    This object is in the DR12Q Supplementary list with $z = 0.622$.

    \subsection{SDSS J225240.37+010958.7}
    {\tt SN\_GAL1} target flag in BOSS. 
    This object is in the DR12Q Supplementary list with $z = 0.533$.

    \subsection{SDSS J233317.38-002303.4}
    {\tt SN\_GAL1} target flag in BOSS. 
    This object is not in DR12Q.

\bsp	
\label{lastpage}
\end{document}